\documentclass[sigconf,authorversion,nonacm]{acmart}

\usepackage[utf8]{inputenc}
\usepackage{xcolor}
\usepackage{hyperref}
\usepackage{graphicx}
\usepackage{subcaption}
\usepackage{array}

\title{\erb{}: Ethics and Society Review of Artificial Intelligence Research}

\author{Michael S. Bernstein}
\affiliation{%
    \institution{Department of Computer Science
\\School of Engineering\\Stanford University}
    \country{}
}
\email{msb@cs.stanford.edu}

\author{Margaret Levi}
\affiliation{
    \institution{Center for Advanced Study in Behavioral Sciences}
    \institution{Department of Political Science\\School of Humanities and Sciences\\Stanford University}
    \country{}
}
\email{mlevi@stanford.edu}

\author{David Magnus}
\affiliation{
    \institution{Department of Pediatrics\\School of Medicine\\Stanford University}
    \country{}
}
\email{dmagnus@stanford.edu}

\author{Betsy Rajala}
\affiliation{
    \institution{Center for Advanced Study in Behavioral Sciences\\Stanford University}
    \country{}
}
\email{betsy.rajala@stanford.edu}

\author{Debra Satz}
\affiliation{
    \institution{Department of Philosophy\\School of Humanities and Sciences\\Stanford University}
    \country{}
}
\email{dsatz@stanford.edu}

\author{Charla Waeiss}
\affiliation{
    \institution{Center for Advanced Study in Behavioral Sciences\\Stanford University}
    \country{}
}

\renewcommand{\shortauthors}{Bernstein, et al.}

\newcommand{\erb}[1]{ESR}

\begin{document}

\begin{abstract}
    Artificial intelligence (AI) research is routinely criticized for its real and potential impacts on society, and we lack adequate institutional responses to this criticism and to the responsibility that it reflects. AI research often falls outside the purview of existing feedback mechanisms such as the Institutional Review Board (IRB), which are designed to evaluate harms to human subjects rather than harms to human society. In response, we have developed the Ethics and Society Review board (\erb{}), a feedback panel that works with researchers to mitigate negative ethical and societal aspects of AI research. The \erb{}'s main insight is to serve as a requirement for funding: researchers cannot receive grant funding from a major AI funding program at our university until the researchers complete the \erb{} process for the proposal. In this article, we describe the \erb{} as we have designed and run it over its first year across 41 proposals. We analyze aggregate \erb{} feedback on these proposals, finding that the panel most commonly identifies issues of harms to minority groups, inclusion of diverse stakeholders in the research plan, dual use, and representation in data. Surveys and interviews of researchers who interacted with the \erb{} found that 58\% felt that it had influenced the design of their research project, 100\% are willing to continue submitting future projects to the \erb{}, and that they sought additional scaffolding for reasoning through ethics and society issues.
\end{abstract}



\maketitle

\begin{figure*}[tb]
    \centering
    \includegraphics[width=1.0\textwidth]{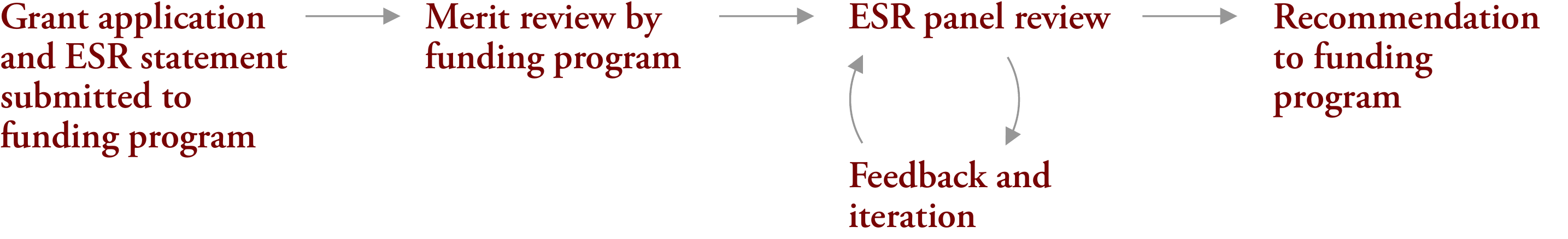}
    \caption{The \erb{} process accepts initial statements from researchers when they submit the grant, then iterates with them prior to funding.}
    \Description[The \erb{} process workflow]{A workflow. Step one: Grant application and \erb{} statement submitted to funding program. Step two: merit review by funding program. Step three: \erb{} panel review. Step four: feedback and iteration. Steps three and four repeat in a loop. Step five: recommendation to funding program.}
    \label{fig:flow}
\end{figure*}   

\section{Introduction}

The observation that artificial intelligence presents societal and ethical risks is a horse that has already left the barn. At this point, that horse has crossed state lines, raced halfway across the continent, attracted international headlines, and landed a shoe company sponsorship. Artificial intelligence~(AI) systems are implicated in generating and propagating disinformation~\cite{bliss2020agenda,hartmann2020next,zellers2019defending}, depressing wages for workers~\cite{alkhatib2019street,mcinnis2016taking,gray2019ghost,lee2018understanding}, perpetuating systemic inequality in policing and the justice system~\cite{buolamwini2018gender,benjamin2019race}, and advancing unequal healthcare outcomes~\cite{obermeyer2019dissecting}. These challenges are often typified by oversights in who is and is not represented in the dataset~\cite{gebru2018datasheets}, who has a seat at the table in the design and deployment of the AI~\cite{zhu2018value}, who is intended to benefit and be harmed by the AI~\cite{costanza2018design}, and what anticipable consequences might arise~\cite{winner1980artifacts,merton1936unanticipated}. AI systems have become embedded into society in ways that reinforce racism, economic disparities, and other social ills~\cite{benjamin2019race,o2016weapons,noble2018algorithms,eubanks2018automating,wachter2017technically,costanza2018design}. While artificial intelligence is not the first academic discipline to wrestle with ethical implications of its work, and the concerns are relevant to computing more broadly, the situation with AI is especially salient today.



Unlike other professions such as law and medicine, computing lacks widely-applied professional ethical and societal review processes. Progress is being made in many dimensions, including algorithmic advances~\cite{dwork2012fairness,barocas2016big,hardt2016equality,agarwal2018reductions}, norm changes~\cite{blodgett2020language}, improved product design processes~\cite{holstein2019improving,rakova2020responsible,veale2018fairness,madaio2020co,gebru2018datasheets,mitchell2019model,wong2021timelines,johnson2021communitypanel}, and both academic and public activism on these issues~\cite{belfield2020activism}. However, ongoing progress requires that everyone involved, not just those who care enough to participate, consider societal impact and ethics in their work. 
Organizing the behavior of an entire group is the role that \textit{institutional structures}---rules, incentives, and processes that apply to all---are designed to play. Examples include journal peer review processes, environmental review of new construction, and the rules of evidence and argumentation followed by courts. In today's university environment, there are few institutional structures to facilitate computing and AI researchers in addressing issues of societal and ethical harm. 

Institutional Review Boards (IRBs) are one possible approach, but IRBs' rules and regulations in the United States~\cite{belmont1979} focus on risks to \textit{human subjects}, not risks to \textit{human society}. The United States Department of Health and Human Services’s definition of human subjects research specifically specifically excludes such review, stating, ``The IRB should not consider possible long-range effects of applying knowledge gained in the research [...] as among those research risks that fall within the purview of its responsibility''~\cite{commonrule}. As a result, when AI research does not directly involve human subjects, many IRBs decline to review research that embeds these harms. Some efforts, such as the Microsoft Research Ethics Review Program established in 2013, take a more expansive view---a view that is unfortunately rare in many research contexts, and one that we hope to amplify with our work. A second approach of recommended processes such as checklists~\cite{madaio2020co}, volunteer drop-ins, and product reviews~\cite{holstein2019improving,rakova2020responsible,veale2018fairness,madaio2020co,gebru2018datasheets,mitchell2019model} all rely on voluntary usage, so while they are valuable, they are limited to those who self-select. A third approach, requirements by academic forums such as NeurIPS~\cite{boyarskaya2020overcoming,abuhamad2020like,nanayakkara2021unpacking} and the Future of Computing Academy~\cite{gibney2018ethics} to add ethics content to research paper submissions, are worthwhile but are only enforced at the end of the research process. In contrast, ethics and societal reflections are best considered at the beginning of the research process, before researchers ossify any decisions in stakeholders, models, data, or evaluation strategy. So, there remains a void for an institutional process in research, which can begin the conversation early on in the project lifecycle and engage with all relevant projects rather than just self-selected opt ins.

We introduce the Ethics and Society Review Board (\erb{}), an institutional process that facilitates researchers in mitigating the negative societal and ethical aspects of AI research by serving as a requirement to access funding: grant funding from a large AI institute at our university is not released until the researchers complete the \erb{} process on the project. Researchers submit a brief \erb{} statement alongside their grant proposal that describes their project's most salient risks to society, to subgroups in society, and globally. This statement articulates the principles the researchers will use to mitigate those risks, and describes how those principles are instantiated in the research design. For example, researchers building a reinforcement learning AI to support long-term student retention might identify that the AI might learn to focus on the learners who it is most likely to be able to retain rather than those most at risk, then describe how they have brought in a collaborator who studies inclusive educational experiences for marginalized communities and will evaluate the system in part for robust transfer across groups. 

The funding program conducts its typical grant merit review, then the \erb{} begins its process on the grants that are recommended for funding (Figure~\ref{fig:flow}). The \erb{} convenes an interdisciplinary panel of faculty representing expertise in technology, society, and ethics to review the proposals and \erb{} statements: our most recent panel included faculty from Anthropology, Communication, Computer Science, History, Management Science \& Engineering, Medicine, Philosophy, Political Science, and Sociology. The \erb{} considers risks and mitigations in the context of possible benefits to society. Its goal is not to eradicate all possible or actual negative impacts---which is often impossible---but to work with the researchers to identify negative impacts and devise reasonable mitigation strategies. It engages in iterative feedback to the researchers, which can include raising new possible risks, helping identify collaborators or stakeholders, conversations, and brainstorming. When the process is complete, the \erb{} submits its recommendation to the funding program, and funds are released to the researchers. We present this process as not the only way to structure such a review---we will certainly iterate its design---but as one such approach that has had some success.

The \erb{} has been active at our university for one year and has reviewed forty one proposals. In this time, all of the large proposals and 29\% of the seed grant proposals iterated with the \erb{} at least once. We conducted an inductive analysis of the \erb{} reviews on thirty five seed grants, observing a focus on issues of harms to minority groups, inclusion of diverse stakeholders in the research plan, dual use, and representation in data amongst the \erb{} panel. We also surveyed and interviewed researchers afterwards to understand their experiences with the \erb{}. 58\% of researchers responded that the \erb{} had influenced the design of their research project, and all of them wished to keep the \erb{} process in place. We engaged researchers outside of the \erb{} faculty and funding program faculty to conduct evaluation interviews with researchers. The interviews identified that researchers desired even more scaffolding and structure to their reflections.

The goal of the \erb{} process is to find a lever that can inject social and ethical reflection early, and to do so with the right amount of friction for researchers. We seek a process where PIs find that their time is well spent, where panelists feel like their time is treated valuably, and where the projects are better for it.
\section{Related Work}
We draw principally on two literatures. In understanding the challenges of creating pro-social AI systems, we draw on the literature of ethics in AI. In designing an institutional response to those challenges, we draw on ethics as it is theorized and practiced in IRB programs and in non-technical disciplines. The \erb{} contributes to these literatures by introducing the first institutional process for considering AI ethics.

\subsection{Ethics and AI}
Negative outcomes stemming from AI practice and research are often raised first through probes that test AIs' behavior across groups. These algorithm audits~\cite{sandvig2014auditing} create controlled, replicable data demonstrating the problem with the algorithm. For example, an audit of computer vision algorithms found that gender recognition algorithms are less accurate with darker skin tones and with women, and especially inaccurate intersectionally with women who have dark skin tones~\cite{buolamwini2018gender}. Other audits have found that social networks return political biases in their search results~\cite{kulshrestha2017quantifying}, image search engines replicate societal biases in occupation searches~\cite{kay2015unequal}, and language models embed age and gender bias~\cite{diaz2018addressing,garg2018word}. Algorithm audits can be a regular practice of government and civil society organizations that use AI~\cite{reisman2018algorithmic}.

Even if an AI system were to meet explicit fairness criteria, it can still become embedded in society in harmful ways~\cite{winner1980artifacts}. For example, facial recognition that is perfectly unbiased can still be used by authoritative states to track down dissidents. And within democratic states, a parade of landmark research documented how AI systems functioning as intended will deny loans, bail, housing resources, or other opportunities to Black people and other groups~\cite{benjamin2019race,o2016weapons,noble2018algorithms,eubanks2018automating,wachter2017technically,costanza2018design}. AIs can also be used to advance problematic agendas as people selectively apply the AI's recommendations~\cite{alkhatib2019street} or undercut its validity~\cite{christin2017algorithms}. Outside of the AI system itself, these systems also rely on a large workforce of unseen and often underpaid contractors who generate the data~\cite{gray2019ghost}.

One thread of work seeks algorithmic solutions to the algorithmic issues. Machine learning training algorithms can be adjusted to enforce mathematical notions of individual~\cite{dwork2012fairness} and group~\cite{barocas2016big,hardt2016equality,agarwal2018reductions} fairness in classification tasks such as recidivism prediction. Other tasks, such as subset selection as seen in search ranking, also feature algorithmic approaches to fairness~\cite{mitchell2020diversity}. A tension in this literature is that there are multiple possible definitions of fairness, and they can generally not be satisfied simultaneously~\cite{corbett2017algorithmic,chouldechova2017fair}. So, existing approaches will often pick a definition of fairness that is appropriate for the violation and task under consideration and develop an approach that enforces it, for example testing counterfactuals~\cite{kusner2017counterfactual}. Typically, these definitions draw from a narrow set of moral theories drawn from deontology and consequentialism~\cite{zoshak2021kant}.

A second thread of work argues that these algorithmic adjustments, while valuable, can never fully address the issue. This literature points out that AI is ultimately an inner loop within broader racist, sexist, classist and otherwise problematic socio-technical systems. As a tool used in unjust systems such as employment and criminal justice, the system will use these tools not to the benefit of minority groups, but against them~\cite{selbst2019fairness,alkhatib2021utopia}. For example, judges can game, critique, or undermine the AI when they dislike its recommendations~\cite{christin2017algorithms}. Moreover, trying to create an interally fair world will ignore the exogenous conditions that gave rise to the current non-fair world~\cite{fazelpour2020algorithmic,bao2021s}. For these reasons and others, one third of people may view a provably fair algorithm as making unfair decisions~\cite{lee2017algorithmic}, and even fair and transprent algorithms may make assumptions about marginalized groups' access to information or their priorities~\cite{robertson2021modeling}. Instead, people apply psychological mechanisms behind procedural justice~\cite{binns2018s}, or simply lose trust in an AI once the concept of fairness becomes salient~\cite{woodruff2018qualitative}.

In response to these issues, researchers have developed guidelines for ethical and pro-social AI development in organizations. These guidelines and principles are extremely numerous: one review identified 63 different sets of ethical AI principles~\cite{mittelstadt2019ai}, and other work concurs~\cite{zeng2018linking,jobin2019global}. These principles share some similarity or even outright inspiration from medical ethics~\cite{mittelstadt2019ai} and are even considered within medical contexts~\cite{char2018implementing}; however, unlike the medical profession, computing has no legal and professional accountability mechanisms, nor formal fiduciary relationship~\cite{mittelstadt2019ai}. High-level principles can be challenging to enact~\cite{holstein2019improving}, so more direct guidance is necessary. Design principles include ensuring human control no matter the level of automation involved~\cite{shneiderman2020human}.
Researchers have instantiated direct guidance as checklists~\cite{madaio2020co} or pre-packaged information about the AI's training dataset~\cite{gebru2018datasheets} and performance across groups~\cite{mitchell2019model}, allowing product managers and engineers to understand the risks up front and work to mitigate them. Breaking out of the product team to engage in participatory methods with communities impacted by the system can help address problems before they arise~\cite{wong2021timelines,johnson2021communitypanel,zhu2018value,wiens2019no}.

Organizations themselves can work to create structural and cultural change to mitigate ethics and societal issues. As organizational theory establishes, no technology directly changes practice or culture---we, and it, are shaped by our organizational contexts~\cite{orlikowski1992duality}. As a result, engineers, product managers, and others within the organization must grapple with a lack of accountability and unknown performance tradeoffs when managing ethical AI~\cite{rakova2020responsible}, not only in the private sector but also in government and civil society~\cite{veale2018fairness}. These underserved organizational influences result in a disconnect between the numerous sets of high-level principles and the on-the-ground decisions they must make in their products~\cite{holstein2019improving}. Further, it remains unclear whose responsibility ethics is within an organization~\cite{metcalf2019owning}, despite a need to audit AI systems throughout the development process~\cite{raji2020closing}. 

Policy solutions can complement organizational solutions by enforcing behaviors across all organizations, not just those that opt in. The IRB system itself arose from a policy requirement for taking public funds. Toward this end, AI researchers have begun organizing activist efforts~\cite{belfield2020activism}. Organizational fixes often come from legislation, and the form this regulation takes can have substantial effects on organizational outcomes~\cite{bamberger2015privacy}. GDPR, for example, has built-in protections for explainability behind AI decisions~\cite{kaminski2020multi}. However, the typical organizational response to regulation is compliance, and compliance efforts are limited in their ability to effect change~\cite{trevino1999managing}. Internal organizational compliance structures do not necessarily lead to more ethical behavior interally, running the risk of serving instead as performative ethics and window dressing~\cite{krawiec2003cosmetic}. Engineers and policymakers may not even share a common definition of which systems constitute AI---those using particular technical approaches (as engineers report), or those approximating cognitive functions (as policymakers report)~\cite{krafft2020defining}.

\subsection{Research ethics review and teaching}
In United States university environments, research ethics conjures notions of the Institutional Review Board (IRB). The IRB arose from the Belmont Report in 1979~\cite{belmont1979}, which itself was a reaction to egregious ethical violations in research such as the Tuskegee syphilis experiment. The Belmont Report identifies three main principles in research: respect for persons, benificence, and justice. These principles are turned into explicit policy via the Common Rule.

Under most current interpretations of the Common Rule by university IRBs, however, issues of societal harm fall outside their purview. The Common Rule articulates that an investigator must be conducting research \textit{about} a human subject---that is, if the research topic is an algorithm rather than human behavior, IRBs such as those at our university view the research as not human subjects research. (And while the definition of ``human subject'' could be expanded to include these risks~\cite{metcalf2016human}, this shift has not in practice become widely adopted yet.) As a result, most AI research falls outside of IRB purview. While calls have been made to subject AI research to regulated ethics review~\cite{calvo2018ai,jordan2019designing}, few organizations require this currently.

\begin{table*}[tb]
    \centering
    \small
    \begin{tabular}{>{\raggedright\arraybackslash}p{2.5cm}>{\raggedright\arraybackslash}p{7.5cm}>{\raggedright\arraybackslash}p{7.5cm}}
        & \textbf{IRB: Institutional Review Board} & \textbf{\erb{}: Ethics and Society Review}  \\
        Focus & Significant risks to human subjects & Significant risks to societies, to groups within those societies, and to the world \\ 
        Requirement & Data collection cannot begin, and funds cannot be spent to pay research participants, until IRB protocol is approved & Grant funding cannot be released by the funding program until the \erb{} process has completed \\ 
        Submission & Specifics of research design, including any procedure followed by research participants and any materials shown to research participants & Goal of research, direct and indirect stakeholders, and higher-level research design. Articulation of principles to mitigate negative outcomes, and description of how those principles are instantiated in the research design. \\ 
        Timing & Regular (e.g., monthly) deadline & Synchronized with grant funding programs \\ 
        Possible outcomes & Approval, return with comments, (rare:) rejection & Approval, return with comments, request synchronous conversation, (rare:) rejection \\ 
        Amendment and renewal & Protocols must be amended if the research design changes, and expire after a fixed period (e.g., 3 years) & Protocols will be examined annually as part of the researcher’s annual grant report to the funding institution \\ 
    \end{tabular}
    \caption{The Institutional Review Board is focused on risks to human subjects, whereas the Ethics and Society Board is focused on risks to society, groups within society, and to the world. To enable engagement with the \erb{} early in the research lifecycle, researchers work with the \erb{} prior to funding being released.}
    \label{tab:irbesb}
\end{table*}

The \erb{} seeks to rectify this situation by indicating that even if policy states that the research does not present significant risk to \textit{human subjects}, it can present significant risk to \textit{human society}. By creating algorithms and systems that may be used as instruments of power, computing researchers bear responsibility for taking the ethical and social dimensions of their work into account in the design and implementation of their research. In our case, there are also some benefits of falling outside of IRB purview: much of an IRB's process is determined by a complex regulatory framework, which can create ethical window dressing rather than true ethical behavior~\cite{trevino1999managing}. Unencumbered by regulatory requirements, the \erb{} can focus on adapting its approach as we learn how to improve. Table~\ref{tab:irbesb} compares the two approaches.

We draw insights and inspiration from the Microsoft Research Ethics Review Program, which applies an anthropological lens to reveal the impacts of research not just on individuals but also on communities and societies. From this position, a much broader swath of research efforts require ethics review. Similar to their effort, we reflect upon a growing need for computing research to consider this broader lens in its research practice. Our position outside of industry entails different institutional opportunities and constraints, motivating in particular our approach of using grant review as a fulcrum enabling us to ask that all researchers consider ethics and societal outcomes. Approaches in industry might identify analogous fulcrums in other forms of resource allocation, for example intern or fulltime position headcount.

Our approach must also acknowledge that computing ethics is complicated by complex questions of consent, publicness, and anonymity~\cite{ethicssurvey2016}. Common practices in AI research such as mining public data remain contested ethical topics. While researchers argue that public data is free to be used, end users may disagree. Anonymization can be broken, putting people at risk~\cite{zimmer2010but,barocas2014big}; participants may be angered when they think researchers are tracking their behavior~\cite{hudson2004go}; public data does not imply consent that the data can be used for research~\cite{barocas2014big}.

The other main lever in universities for computing ethics is its educational work. Ethics courses in computing cover issues such as law and policy, surveillance, inequality and justice, and often, AI and algorithms~\cite{fiesler2020}. Alternatively, this content can be integrated broadly across the curriculum~\cite{grosz2019,ko2020}---though it rarely arises in machine learning courses currently~\cite{saltz2019}. Educators are developing many different techniques for teaching this material in traditional courses~\cite{skirpan2018}, including mock trials~\cite{canosa2008}, science fiction~\cite{burton2018}, and training on one's own personal data~\cite{register2020}.
\section{The \erb{} Process}

\begin{figure}[tb]
    \centering
    \includegraphics[width=1.0\columnwidth]{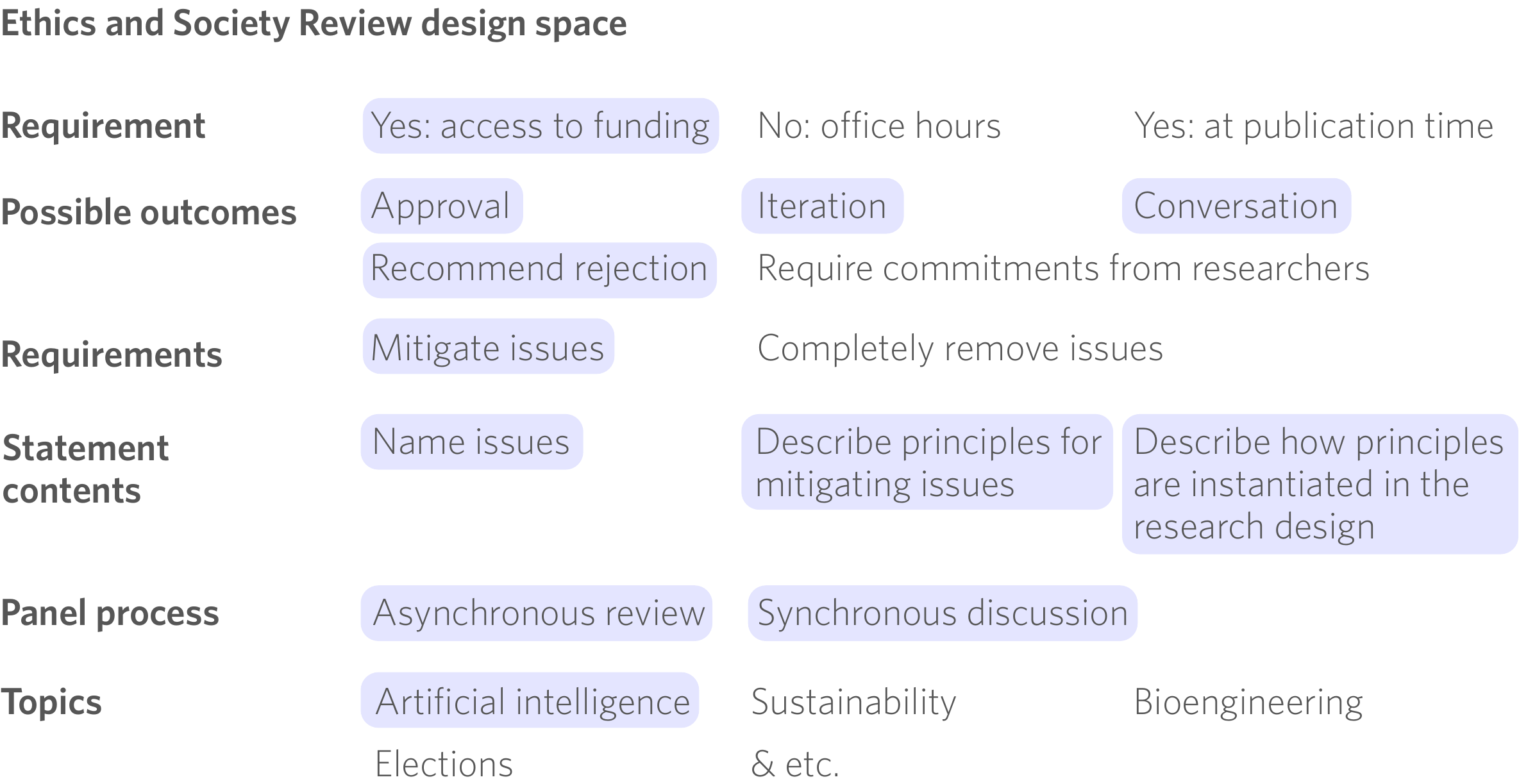}
    \caption{There are many alternative instantiations of an \erb{}. Here we summarize some of our major decision points.}
    \Description[Design space with highlighted options]{Design axes in columns, followed by options for each axis and highlighted options. Requirement axis: yes, access to funding (highlighted); no, office hours; yes, at publication time. Possible outcomes axis: Approval (highlighted), Iteration (highlighted), Conversation (highlighted), Recommend rejection (highlighted), Require commitments from researchers. Requirements axis: mitigate issues (highlighted), completely remove issues. Statement contents: name issues (highlighted), describe principles for mitigating issues (highlighted), describe how principles are instantiated in the research design (highlighted). Panel process axis: asynchronous review (highlighted), synchronous discussion (highlighted). Topics axis: artificial intelligence (highlighted), sustainability, bioengineering, elections, etc.}
    \label{fig:designspace}
\end{figure}

We designed the \erb{} to hook into internal funding grant review at our university. In this section, we describe how we designed the process (Figure~\ref{fig:flow}), and our motivation in making each of these design decisions (Figure~\ref{fig:designspace}).

\subsection{Partnership with funding program}
The most critical institutional feature of the \erb{} is the collaboration with a funding program. This collaboration enables completion of the \erb{} process to trigger release of funds. Funding is a rare moment of institutional leverage in the university: while most AI research proceeds without IRB review at our university, researchers are often in search of funding.

We partnered with a cross-departmental institute at our university that runs funding competitions with both (1)~a large, multi-PI grant competition with a small number of projects receiving substantial funding, and (2)~a smaller seed grant competition with many projects receiving less funding. Working with our team, the institute (program) added a requirement for an \erb{} statement for each grant submission. The program performed its typical merit review on all grant submissions, and sent the \erb{} the proposals that they were interested in funding. The \erb{} then performed its own internal process on those proposals, and reported its outcome and recommendation to the program.

This collaboration with the funding program requires several points of collaboration from them. First and most obviously, the program is allowing an external group some influence over its grant decisionmaking. In rare situations where the \erb{} process has failed to yield a satisfying outcome, the \erb{} might recommend not funding a grant. In our case, the funding program wished to retain final decision rights after receiving the \erb{}'s recommendation. Second, the program needs to build additional time into its grant review process to allow the \erb{} to perform its function, and agree that the \erb{} may reach out to the grantees for conversation and feedback before they have heard official funding notification from the program. Third, the program needs to provide reasonable estimates ahead of time of the volume and types of submissions that it expects to receive, so that the \erb{} can form a panel.

\subsection{Submission: The \erb{} statement}

For researchers, their first exposure to the \erb{} is by writing an \erb{} statement about their project. The \erb{} statement is an explicit space for the researchers to reflect on the ethics and social issues with their project, and to explain what they are doing to mitigate those issues. 

As a prompt, researchers are asked to detail the ethical challenges and possible negative societal impacts of the proposed research. Then, they are asked to explain how they will mitigate those impacts. They are instructed to, for example, consider autonomy and consent from those they are getting data from, who is and is not represented in the training and test data, who will be impacted by the technology once it leaves the lab, and whether the approach is importing or creating bias.

To aid researchers in structuring their thinking, the \erb{} statement prompt asks researchers to organize their statement into two parts, quoted below, with particular emphasis on the inclusion of the second element:
\begin{enumerate}
    \item \textit{Describe the ethical challenges and societal risks.} What are the most important challenges you face with your project? Consider the following three groups in your response: (1)~Society: the society targeted by the research, considered as a whole (e.g., American society); (2)~Groups within society: risks are not distributed equally amongst a society, and marginalized groups may be especially vulnerable  (e.g., Black individuals, LGTBQ+ individuals); (3)~Global: impacts on the world as a whole, or on societies that are not directly targeted by the research but that may be impacted by the research (e.g., potential abuse in developing regions).
    \item \textit{Articulate general principles that researchers in your field should use to eliminate or mitigate these issues, and translate those principles into specific design decisions you are making in your research.} Think of what happens when someone else builds on your work or a company outside of your control decides to replicate it. What principles should others in this field follow when faced with similar tradeoffs? How does your proposal instantiate those general principles? If your research team does not currently have required expertise or perspectives represented, how will you obtain them?
\end{enumerate}

The instructions include examples. For example, if (1)~includes a risk that a new healthcare algorithm is biased against Black members of society, a researcher might propose in (2)~that all such algorithms must be audited against risks for under-represented groups, then describe how they will collect data to audit the algorithm against bias for Blacks, Latinx, Native American, and other under-represented groups. In future iterations, based on researcher feedback, we also plan to include prompts suggesting common categories of issues that arise in \erb{} processes.

The minimum \erb{} statement length ranges from one page to several pages depending on the project topic, size, and funding level. It is not the final say on the matter; it is the starting point of the conversation.

\subsection{Panel Review and Feedback}

The funding program next performs its grant merit review process, and selects proposals that it would like to fund. The proposals and their \erb{} statements are then forwarded on to the \erb{} for feedback. The \erb{}'s goal is not to filter out projects with any modicum of risk---instead, when possible, the goal is to aid the researchers in identifying appropriate mitigation plans.

The \erb{} faculty panel is composed to bring together diverse intellectual perspectives on society, ethics, and technology. Our panel thus far represents faculty from the humanities, social sciences, engineering, and medicine and life sciences. Their departments at our institution include Anthropology, Communication, Computer Science, History, Management Science \& Engineering, Medicine, Philosophy, Political Science, and Sociology. Their interests include algorithms and society, gender, race, ethics, history of technology, collective action, medical anthropology, moral beliefs, medical ethics, social networks, AI, robotics, and human-computer interaction. Many other disciplines and identities can and should be included as well.

Each proposal is assigned to at least two panel members, one representing the broad field of inquiry of the proposal (e.g., medicine, engineering, social science), and one representing a complementary perspective. A few chairs take on the role of chairs in facilitating the feedback process, overseeing the feedback process for individual proposals.

Each panelist reads their assigned proposals and the associated \erb{} statements, then responds with two pieces of information per proposal:
\begin{enumerate}
    \item Level of concern in the ethical and societal implications of the project, with 1=``no concern'' and 5=``very concerned''. This score is not shared with the researchers, but is used internally to prioritize discussions and establish cutoffs for requesting iteration.
    \item Open ended feedback. Panelists are asked to focus on the questions in the prompt: (a)~What are the ethical challenges and negative societal impacts? (b) Did they identify principles or strategies to mitigate those challenges and impacts? (c) Did they instantiate those principles or strategies in their research design?
\end{enumerate}

To help with training, panelists are provided with example past proposals and the \erb{} responses for them. The \erb{} chairs also organize open video calls for panelists to join and discuss if desired.

The \erb{} panel then meets synchronously to discuss particularly controversial or challenging projects. The chairs flag projects where the quantitative score, the level of concern, is highest. The panelists then describe the proposal and their reactions to it, and other panelists provide their suggestions and input. This meeting provides an opportunity for the panel to meet, set norms, and learn from each other. At the conclusion of the meeting, the chairs and panelists decide on which proposals will be recommended for funding directly, and which will be asked to engage in further conversation and feedback. Finally, panelists have a short period of time to edit their comments before they are sent out to the researchers.

\subsection{Iteration and approval}
The goal of the iterative \erb{} process is to help researchers identify effective responses to the risks proposed by their projects. While the \erb{} can ultimately recommend not funding a project, we view that outcome as a last resort only after sustained iteration and communication with the researchers about their project does not result in a feasible path forward.

The first step in this iterative process is when the researchers receive their feedback from the \erb{} panel. All researchers recieve the freetext feedback provided by the two panelists. A subset are told that the \erb{} has completed its process on the projects and it will recommend the project for funding, though it welcomes further discussion if the researchers desire. Typically, these projects have low levels of concern from the panel.

The second subset of proposals decided upon by the \erb{} committee are asked to respond to the \erb{}'s feedback. The \erb{} chairs make themselves available for conversation and consultation with the researchers. When the reseachers respond, the response is passed back to the relevant panelists, who provide their thoughts and recommendation to the \erb{} chairs. The \erb{} chairs then draft a response to the researchers representing the \erb{}'s thoughts and their own assessment, and send it back to the researchers. Future iterations remain on email if the discussion is converging, or can switch to a synchronous meeting if not. The \erb{} chairs become the point of contact for the researchers follwing the first round of feedback in order to avoid jeopardizing colleague relations (the ERB first round feedback is authored anonymously), and to help facilitate the most challenging projects.

\subsection{Participatory process and evolution}

We have endeavored to engage in a participatory process in the design of the \erb{}, including discussions with researchers, with the funding agency, and with the \erb{} panel. An obvious limitation and point of departure for next steps is to include community stakeholders. Here, we report some of the most salient points of evolution thus far.

A substantial amount of normsetting needs to occur across disciplinary boundaries. In the \erb{} panel meeting, for example, opinions ranged on how aggressively the panel should be calling out issues or recommending rejection. There was also discussion about the conditions under which methodological issues in the proposal are considered ethical issues---the field of sociology, for example, mentions methodological issues as potential ethical issues because incorrect methodology undermines trust in the field and promulgates incorrect conclusions that impact public policy decisions.

One common misunderstanding in early iterations of our approach was that researchers' \erb{} statements named possible issues, but then proceeded to minimize those issues' likelihood in order to downplay the risks of the project. 
This approach was counter to the \erb{}'s goals of trying to actively mitigate negative outcomes. So, we included the second requirement to explicitly name strategies for mitigating the outcome and implementing them, rather than simply arguing it is unlikely.

A second common misunderstanding we observed in \erb{} statements is that researchers focus on issues that should be considered by an IRB, not an \erb{}. For example, we received several \erb{} statements discussing risks to research participants (e.g., how to prevent a robotic assistive device from harming people in its laboratory user studies). In these instances, the \erb{} engages further with the researchers in the feedback and iteration rounds. We have iterated our instructions and feedback to emphasize what does and does not belong in the \erb{} vs. IRB.

We have found that \erb{} panelists' natural inclination, given the academic environment, is to act as reviewers rather than coaches in their feedback. This remains an open challenge, as the \erb{}'s purported goal, especially in initial rounds of feedback, is to help researchers improve their project rather than act as gatekeepers. Additional discussion is needed around what our goals can and should be.
\section{Deployment and Evaluation}     

The \erb{} has been active at our university for about a year. In that time, it has facilitated feedback on 41 grants: six large grants and thirty five seed grants. All six of the large grants (100\%) and ten of the seed grants (29\%) were asked to iterate based on the \erb{}'s feedback, of which three (9\%) iterated multiple times. All were eventually supported by the \erb{}, not as risk-free but as having appropriate mitigation plans. 

In this section, we describe our deployment of the \erb{} thus far, and evaluate its effectiveness: Does it provide a useful scaffold? Which elements are funtioning well, and which can be improved? We answer these questions through a combination of surveys and interviews of researchers who participated, as well as quantitative analysis of the \erb{} panelists' ratings and feedback. Our results suggest that researchers found the \erb{} to provide a useful scaffold, and wanted it to go even farther in providing scaffolding for reasoning through their own projects.

\subsection{Method}
We surveyed the researchers who engaged with the \erb{}'s seed grant process to gather feedback on their perceptions of the \erb{}. The survey was sent to the lead researcher of the 35 grants that the funding program wanted to fund and asked the \erb{} to review. The survey investigated questions of researchers' other involvement in ethics issues in their research, the level of influence that the \erb{} feedback had on the project, the aspects of the process that the researchers found most helpful and least helpful, and opinions on whether the \erb{} can help mitigate negative outcomes. The survey and follow-up interviews were both covered by an IRB-approved consent process. Researchers from twenty three projects responded to the survey and twelve did not, a 66\% response rate.

Following the survey, we invited the lead researcher to join a semi-structured interview of 30--60 minutes in length. These interviews were conducted from December 14, 2020 to February 5, 2021. Sixteen grantees from fifteen different projects participated in the interviews. The interview protocol is available in the Appendix. Researchers from fifteen projects participated in the interview, a 43\% interview rate amongst all projects. Five projects represented in the interviews underwent at least one revision during the \erb{} process; the remaining ten projects passed the \erb{} process without revisions. The interviews represented projects with focuses in engineering~(7), social science~(4), earth science~(2), and medicine~(2).

All analyses in this section are exploratory, and none are explicit hypothesis testing; thus no hypotheses or designs were preregistered. Furthermore, because interviews and surveys took place after grantees’ participation in the \erb{} process, we cannot determine participants’ baseline attitudes and behaviors toward research ethics prior to interacting with the \erb{}. Therefore, these results could represent an attenuated treatment effect from the \erb{}, participants’ return to their baselines, or outcomes that are endogenous to other factors influencing participants’ attitudes and behaviors toward research ethics. In addition, a few participants indicated that they chose to apply for the grant program because of the added \erb{} process. Consequently, some participants could be more inclined toward thinking about research ethics, or more positively inclined toward the \erb{}, than those who did not participate. 

Data from \erb{} panelists' reviews were analyzed via a consent process after the \erb{} process was complete. Fourteen panelists consented, and one panelist declined. Anonymized, aggregate numerical review scores from the funding program were shared by the funding program. Because an author of the paper serves as a Dean and holds a supervisory relationship to some researchers and panelists, that coauthor did not analyze raw or anonymized data in this section, mitigating the possibility of a conflict of interest in consent.

\subsection{Case studies: the \erb{} in practice}

We begin with a trio of case studies, to make concrete the forms of interaction that the \erb{} had with grant proposals. We withhold details of the grants where reasonable to do so.

\subsubsection{AI Teaching Agents: displacement, representation, and optimization targets}
One of the larger grants reviewed by the \erb{}, which included six faculty crossing Computer Science, Education, and Psychology, focused on the creation of a new generation of AI tutoring systems. The systems would be evaluated through field deployments online and in brick-and-mortar schools.

The \erb{} focused its initial feedback on three questions. First, the panel wanted to hear more about whether the research would be used as a justification to displace teachers' jobs. In their \erb{} statement, the researchers had stated, ``we design our tools to force-multiply teachers so that they can have more impact, and to only provide fully autonomous education in contexts where there is no human option''. The panelists asked the researchers to clarify what, concretely, in the proposed system would make it more useful as a force multiplier than as a replacement. 

The second discussion point was: who will be represented in the data? It would be important to make sure that the resulting model is not just useful to the communities already have the capacity to benefit~\cite{toyama2015geek}. The panel asked the researchers to clarify whether background diversity would be reflected in the dataset collection, and how that diversity would be recruited.

The final question was about what the models would learn: depending on how it is implemented, optimizing for long-term engagement could cause the algorithm to increase its reward most effectively by focusing on the learners who it is most likely to be able to retain. This would mean it would bias against groups of learners who drop out at high rates — traditionally groups of low socio-economic status students — since the policy would learn that they are unlikely to be retained, so it should focus its effort on others. The panelists asked the researchers to propose how they would guard against this?

The researchers then responded to each of the three points. With respect to job displacement, the researchers pointed to research identifying teachers as the least likely jobs to be displaced by AI, and committed to a specific design constraint on their systems that they assume a teacher copresent in the room alongside the student. With respect to representation, the team committed to the process being led by one of their team members who specializes in the design of
inclusive educational experiences for marginalized communities, and further committed to explicit transfer testing with diverse groups.  Finally, with respect to undesired outcomes of the optimization criteria, the team stated that answering those questions was an explicit goal of the research, and committed to engaging with experts in educational equity throughout the project.

The panelists felt that the responses responded to the questions and committed to specific mitigation strategies, and so passed the proposal.

\subsubsection{Remote sensing AI: malicious actors}
A seed grant featuring faculty from Earth Systems Science and Computer Science proposed an approach to creating remote sensing models to be used in sustainability applications. In their \erb{} statement, the researchers pointed out two possible threats: that the models could be adopted by intelligence and military organizations in authoritarian states to enhance mass surveillance, and that the representations they developed might perform differently in different parts of the world. The researchers pointed out that the mass surveillance applications were less likely to arise from their research because such state actors already had resources to collect the large-scale data needed to train such models, making the incremental risk of this project more minor; and the researchers committed to focusing their models solely on sustainability applications. With respect to performance imbalances, the team committed to benchmarking their models globally, and specifically focusing on Africa, to challenge the status quo of similar models focusing on America.

The \erb{} panelists, in their responses, focused on a theme of how to guarantee non-malicious use. They pointed out that naming the possibility of malicious use itself does not carry teeth, and that the statement relies on an assumption that other datasets are more attractive to malicious actors than the datasets and models produced in this research. The panelists felt that a paired forensic model was a good commitment, and suggested that the researchers gather data on how people in each area are surveyed currently and what impact this shift might give rise to. 

Ultimately, the panelists felt that this feedback was sufficient and did not request the panelists to respond.

\subsubsection{Stress sensing: privacy}
One of the funded seed grants, proposed by faculty in Medicine and Electrical Engineering, focused on noninvasive stress sensing at work. In their \erb{} statement, the researchers expressed an awareness of possible concerns about repressive governments and oppressive employers as two examples of groups that would use this technology to surveil and potentially harm. The statement did not provide principles that researchers should use to mitigate that risk, or concrete instantiations of those principles in their research, instead stating that the team would take user agency and privacy seriously.

The panelists asked the researchers to iterate on their statement and explain specific design features that they would be including to address the risks. In addition, one panelist pointed the researchers to prior work on commiting to returning information about stress levels to the participants at the conclusion of the study and on an ongoing basis if appropriate.

The researchers and cognizant \erb{} chair in charge of that proposal then met to discuss the feedback. Following the discussion, the researchers wrote their response to the committee feedback, which focused on how the researchers would protect data, but did not discuss what should be done early in the research to account for risks once the technology leaves the lab and is commercialized. In response, the researchers committed to explaining the privacy-preserving aspects of their design, and the importance of those aspects, in all papers arising from this research---essentially using their bully pulpit at our institution to push for norms in how this technology is developed and deployed. The \erb{} felt that the combination of this commitment with the conversations held with the researchers was appropriate to the level of risk, and agreed to recommend the proposal.

\subsection{Main benefit of the \erb{}: scaffolding}

\begin{figure}[tb]
    \centering
    \includegraphics[width=\columnwidth]{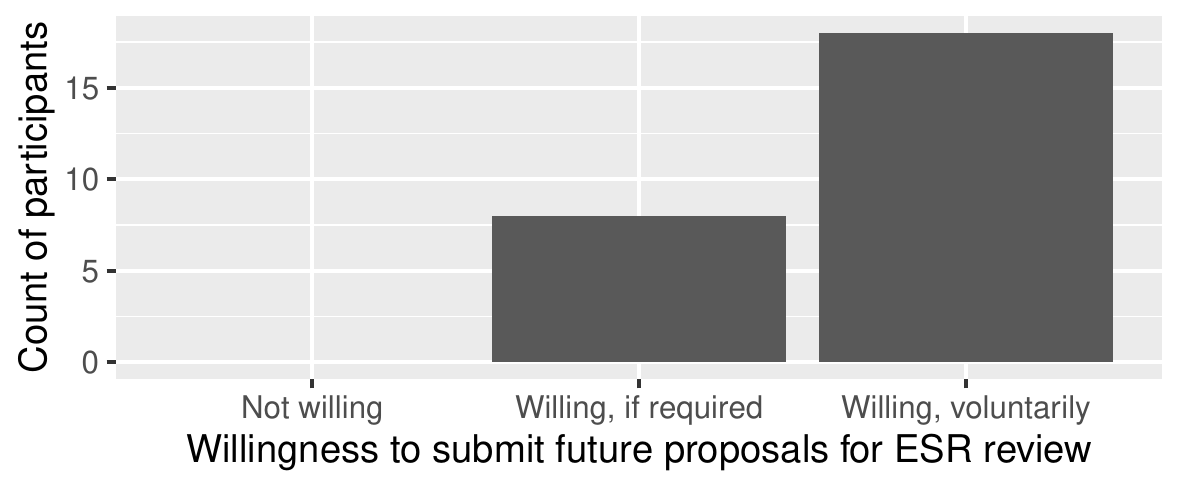}
    \caption{All participants were willing to engage in the \erb{} process again.}
    \Description[Histogram of how willing participants were to participate again]{Histogram with X axis, ``Willingness to submit future proposals for ESR review''. Roughly eighteen participants responded ``willing, voluntarily'', roughly eight participants responded ``willing, if required'', and zero participants responded ``not willing''.}
    \label{fig:willingness}
\end{figure}

Overall, researchers were positive on the \erb{} experience and wished to continue it. The survey asked participants whether they would submit to the \erb{} again. All were willing (Figure~\ref{fig:willingness}). Stratifying the responses by whether grants were asked to iterate with the \erb{}: amongst those who did not iterate with the \erb{}, 37\% said they would only do it if required and the rest (63\%) said they would do it voluntarily; amongst those who iterated with the \erb{}, all said they would do it voluntarily.

\begin{figure}[tb]
    \centering
    \includegraphics[width=\linewidth]{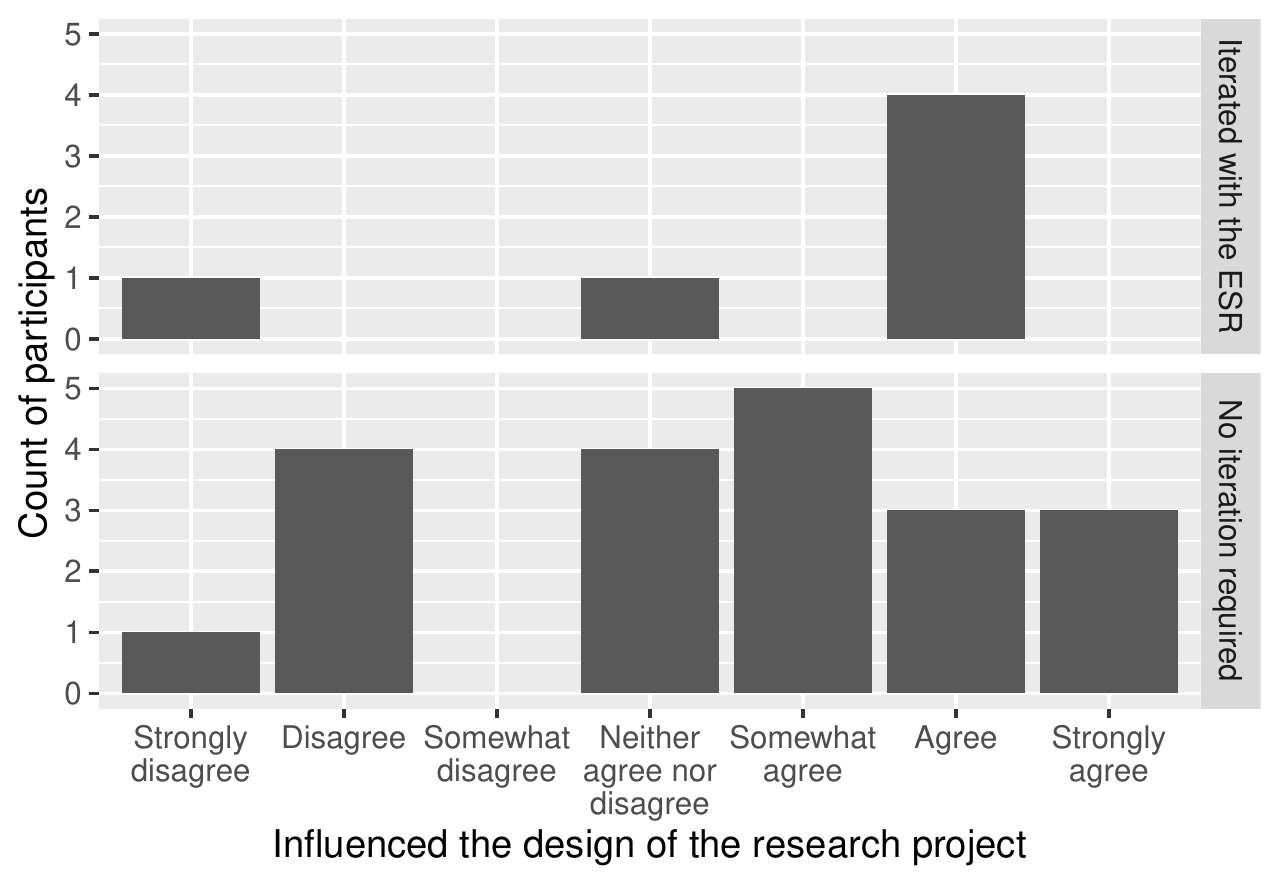}
    \caption{67\% of researchers who iterated with the \erb{}, and 58\% of all researchers, felt that the \erb{} process had influenced the design of their project.}
    \Description[Histogram of responses to question ``influenced the design of the research project'']{Histogram, with X axis labeled ``influenced the design of the research project''. Graph is split into two rows. In the top row, labeled ``iterated with the ESR'', four respondents replied ``agree'' and one each said ``strongly disagree'' and ``neither agree nor disagree''. In the bottom row, labeled ``No iteration required'', five respondents replied ``somewhat agree'', four replied ``disagree'' and ``neither agree nor disagree'', three responded ``agree'' and ``strongly agree'', and one responded ``strongly disagree''.}
    \label{fig:design}
\end{figure}

58\% of the self-reported responses indicated that the \erb{} process had influenced the design of the research project (Figure~\ref{fig:design}). Most projects did not iterate with the \erb{}, so the parts of the process they experienced were the writing of the \erb{} statement and reading the \erb{} panel's feedback. Among those who iterated with the \erb{}, 67\% indicated that the \erb{} process had influenced their design.

Nearly all interviewees reported that the \erb{} process encouraged to think more deeply about the broader implications of their research. Eight participants said that the \erb{} process raised new issues for them to think about. For six others, while the process didn’t raise new issues, it encouraged them to contemplate even further the ethical implications they had already been considering.
In addition to encouraging respondents to think more deeply about the societal and ethical implications of their work, both the forcing function of the \erb{} statement and the panel's feedback led participants to discuss the issues with others. For a couple of respondents, these conversations also revealed new issues to consider.

Overall, the \erb{} process also appeared to raise the consciousness of some researchers to engage more seriously with research ethics going forward:
\begin{quote}
    I don't consider myself an expert, by all means, but I'm definitely very interested [in research ethics] and I think the [\erb{}] has started that spark in me. I’m very interested in doing it -- like, actively pursuing it. But I still need to practice and learn ... theory ... it's necessary for these things, so I'm still in the learning phase, but I'm very motivated. Let’s play. -~Researcher, medicine 
\end{quote}

\begin{quote}
    The [\erb{} statement] requirement … led me to engage with my co-PI … because, as a psychologist, I ... wasn't aware of some of the potential ethical implications that this … AI work may have, and it helped me to engage with my co-PI as part of this requirement. -~Researcher, social science
\end{quote}

In terms of participants’ project designs, four interviewees could identify concrete components of their research that they changed as a result of the \erb{} process. Two participants discussed significant changes. One even changed the guiding research question of their work: 
\begin{quote}
    In fact, we might flip our whole research approach to being about privacy. And one application is well-being and buildings. So it's almost like … the leading research questions are around how do you create some system like this with privacy-enhancing measures and, in our case, we're applying it to buildings. 
    
    [The] pretty strong reaction from the [\erb{} made] us rethink, to lead with … privacy. We really just want buildings to be spaces that people flourish in and we need to do it in some way that's going to be the most privacy-preserving [as] possible … We don't have answers yet, but … it's definitely helped us think about a better way to approach the research, how we're doing it and how we're talking about it. -~Researcher, engineering 
\end{quote}

For two other interviewees, one noted changes to how they engaged with local experts. The other modified their approach to privacy and attending to the dual-use nature of their project. Six other respondents were at such early points in the research process that, while they couldn’t identify concrete changes the \erb{} process yielded, they anticipated their experience with the \erb{} had shaped and would continue to influence the development of their projects. One participant referred to this as “ethics by design.” 

Of the remaining 6 interview participants, 5 did not believe the \erb{} led to any changes in their project design. One respondent believed it could have, but the issues raised by the panelists related to data collection, which the respondent had already completed and could not modify. This is not surprising, given that projects with low risk were passed by the \erb{} panel with less feedback.




Writing the statement served as a commitment mechanism for some respondents. It not only required that participants commit their thoughts to paper about the ethical implications of their work and any necessary mitigation strategies, but the statement would also be reviewed by the \erb{} panel:
\begin{quote}
    I think just the brainstorming aspect of being like, “Oh yeah, there's an [\erb{}], they're going to think about our project. We don't want to be embarrassed by not having thought about obvious downstream harms of our thing. So let's just sit down and spend two hours ... coming up with bad ways of using this technology.” -~Researcher, engineering 
\end{quote}

At the same time, participants responded favorably to the amount of writing required for the statement (about one page). When prompted, no respondents felt the writing requirement needed to be reduced. In fact, one participant expressed concern that reducing the length limit of the statement would prove more cumbersome, as it would require them to synthesize a response into such a short space. Others felt similarly that extending the length of the statement would also be too cumbersome. Although respondents were eager to share additional ideas and elements to include in the \erb{} process, nearly all commented on time constraints that would prevent them from participating if the process were more burdensome.

Most researchers had little prior experience with ethics and society reflections. On one hand, most survey participants reported engaging in conversations about ethics prior to the \erb{} process. 8\% reported having such conversations within one day prior to the survey, 27\% within one week, 46\% within one month, and 19\% longer or could not recall. However, nearly all interview participants (92\%) had not previously engaged in a structured ethics review outside of the \erb{}: one named the NeurIPS submission requirement, and another an involvement with an on-campus ethics center. And although a majority of interviewees (10) mentioned engaging with research ethics frequently or increasingly, when prompted about the nature of those research ethics, only 3 respondents --- one from each area of study, except engineering --- cited ethical issues beyond the scope of the IRB, and four others discussed research ethics in the context of the IRB. Another mentioned the ethical considerations related to identifying sources of research funding. So, the \erb{} appears to be supporting a unique need for these researchers.

Ultimately, researchers felt that the \erb{} process made it less likely that their project would misstep and wind up in the public eye for the wrong reason. 73\% of survey respondents agreed that the \erb{} reduced the probability of public criticism of their project, 
with 100\% of those who iterated with the \erb{} agreeing.

\subsection{Concerns raised in \erb{} feedback}
We first sought to analyze: what risks do the researchers writing proposals see in their project, and what risks does the \erb{} bring up in response?

One of the authors conducted open coding across all of the \erb{} statements and panelist responses using a grounded theory method to develop a set of 14 codes of themes brought up. These codes and their definitions are included in Table~\ref{tab:codes}. A second author independently coded a subset of statements and panelist responses to test replicability; inter-rater reliability via Cohen's kappa averaged 0.96 per theme, with a range of 0.83 - 1.

\begin{table*}[tb]
    \centering
    \small
    \begin{tabular}{>{\raggedright\arraybackslash}p{2.25cm}>{\raggedright\arraybackslash}p{2.25cm}>{\raggedright\arraybackslash}p{2.25cm}>{\raggedright\arraybackslash}p{10cm}}
        \textbf{Theme} & \textbf{Researcher statement frequency (N=35~proposals)} & \textbf{Panelist response frequency (N=35~proposals)} & \textbf{Refers to issues that pertain to\ldots}  \\
        Representativeness & 18 & 6 & Any risks or concerns that arise from insufficient or unequal representation of data, participants, or intended user population (e.g., excluding international or low-income students in a study of student wellbeing) \\
        IRB purview & 14 & 6 & Any risks or concerns regarding the research that fall under IRB purview (e.g., participant consent, data security, etc.) \\
        Diverse design and deployment & 13 & 8 & Incorporating relevant stakeholders and diverse perspectives in the project design and deployment processes (e.g., consulting with parents who have been historically disadvantaged to develop fairer school choice mechanisms) \\        
        Dual use & 10 & 8 & Any risks or concerns that arise due to the technology being co-opted for nefarious purposes or by motivated actors (e.g., an authoritarian government employed mass surveillance methods) \\
        Harms to society & 10 & 5 & Potential harms to any population that could arise following from the research (e.g., job loss due to automation) \\
        Harms to subgroups & 7 & 11 & Potential harms to specific subgroup populations that could arise following from the research (e.g., technical barriers to using an AI that is prohibitive to poorer populations) \\ 
        Privacy & 4 & 1 & Any risks or concerns related to general expectations of privacy or control over personally identifiable information (e.g., consequences of mass surveillance systems for individuals' control over their information) \\ 
        Research transparency & 3 & 0 & Sufficiently and accessibly providing information such that others can understand and effectively employ the research, where appropriate (e.g., training modules for interpreting an AI model)\\ 
        Accountability & 2 & 2 & Questions of assigning responsibility or holding actors accountable for potential harms that may arise (e.g., how to assign responsibility for a mistake when AI is involved) \\ 
        Other & 2 & 3 & Other issues not covered above (e.g., intellectual property concerns) \\ 
        Tool or user error & 2 & 4 & Any risks or concerns that arise from tool/model malfunction or user error (e.g., human misinterpretation of an AI model in decision-making)\\ 
        Collaborator & 1 & 1 & Any risks or concerns that specifically relate to a collaborator on the research project (e.g., whether a collaborator could credibly commit to a project on inclusivity when their platform was notorious for exclusive and harmful behavior) \\
        Methods and merit & 1 & 2 & Any risks or concerns reserved for methods and merit reviews of the grant proposal (e.g., whether model specifications are appropriate for the goals of the research) \\
        Publicness & 0 & 2 & Questions of using publicly available data for research when those that generated the data are unaware of researchers’ intended use of their data (e.g., use of Twitter data without obtaining express consent from affected Twitter users) \\
        
    \end{tabular}
    \caption{The researchers, in their \erb{} statements, were most likely to raise issues of representativiness. The panelists, in their feedback, were most likely to raise issues regarding harms to subgroups. Both researchers and panelists also commonly focused on diverse design and deployment, dual use concerns, harms to society, and issues pertaining to IRB purview.}
    \label{tab:codes}
\end{table*}

For researchers, each risk or issue identified in the \erb{} statement was coded along the above risk themes. Although some issues could pertain to multiple risk areas, we only coded them along the risk area(s) identified by the PIs. For example, one statement discussed the risk that a facial recognition system may make mistakes and suggested a mitigation strategy that allowed users to report such errors. Therefore, this was coded as a tool / user error risk. Had the PIs also discussed whether specific subgroups were more likely to be misidentified by the system than others and the consequences of such widespread errors for those subgroups, this would have also been coded as a harms to subgroups risk. Constraining the coding in this manner ensures that we adequately capture the ethical issues PIs found salient in their work, rather than imparting additional ethical lenses they may not have employed. The only exception to this is the coding for IRB purview risks. Due to the prevalence of issues that PIs and panelists both raised regarding participant consent, data privacy, and potential harms to participants, we created the IRB purview category to capture these.

Due to the nature of the feedback provided by panelists, we first coded the actions that a panelist took in their feedback. These include: identifying a new risk, providing specific mitigation strategies, proposing potential collaborators to the PIs, referring the PIs to specific previous work on a relevant ethical issue, and deepening the conversation. When a panelist raised a new ethical risk or issue that PIs had not explicitly identified in their \erb{} statement, we coded such issues along the same risk themes in Table~\ref{tab:codes}. An element of panelists’ feedback was coded as deepening the conversation if it generally engaged with the researchers’ \erb{} statement, but did not constitute one of the other actions. For example, questions that probed researchers to think more deeply about an ethical issue they raised in their \erb{} statement would be coded as deepening the conversation. Finally, we also coded instances of panelists’ feedback that neither took any of the actions outlined above nor mentioned any new ethical issues.

Altogether, the issues raised most frequently by PIs and panelists fell into similar categories (Table~\ref{tab:codes}). For PIs, the most popular issues mentioned in \erb{} proposals were (in order of frequency): representativeness (18), issues that fell under IRB purview (14), diverse design and deployment (13), dual-use (10), and harms to society (10), followed closely by harms to subgroups (7). For panelists, risks related to harms to subgroups (11) were the most frequently cited, followed by diverse design and deployment (8), dual-use (8), representativeness (6) and issues that fell under IRB purview (6). Concerns relating to harms to society (5) were next.

The high frequency of IRB-related risks in both the \erb{} statement and -- to a lesser extent -- panelists' feedback highlights an important area for improvement going forward, as both researchers and panelists raised issues that the IRB should cover, including how data is protected, and risks to direct research participants. Even in cases where PIs did not discuss IRB-relevant information in the \erb{} statement, panelists often requested additional information or expressed concern that the \erb{} statement did not address issues related to IRB concerns, like participant consent or data privacy. Our hope is that better training and norm-setting could alleviate this issue.

Unlike the overlap in issues that the IRB would review, PIs and panelists rarely raised issues related to those that would fall under a traditional methods and merit review (PIs: 1; Panelists: 2). Nevertheless, this highlights another area that future iterations of the \erb{} will seek to address. 

On the whole, the \erb{} statement and panelists' feedback revealed that both PIs and panelists were grappling with many ethical issues that would not have otherwise been addressed. It also appears that panelists broadened the ethical scope for PIs in a responsive manner. For example, as a majority of panelists already discussed issues of representativeness in their ethics statements, panelists were less likely to raise those issues. Instead, panelists raised additional issues in those cases, like harms to subgroups and dual-use concerns. 

Of the 35 projects reviewed during this cycle, panelists on 26 of those projects raised new ethical issues that the PIs had not discussed in the \erb{} statements.
Panelists raised new ethical issues for 18 of the 25 projects that did not require iteration with the \erb{}. This result illustrates that panelists’ feedback was not constrained by the content of the \erb{} statements and served to broaden the ethical scope that PIs used in their projects. 

Panelists pointed out ethical issues that researchers hadn’t discussed in their \erb{} statement for all but 2 proposals that iterated with the \erb{} (80\%). For those two proposals, panelists’ feedback concentrated on requesting further details from researchers and offering mitigation strategies for some of the ethical issues that were unresolved by the discussion in the \erb{} statement. Only for 4 projects did panelists constrain their feedback to what was included in the \erb{} statement. In one of these instances, a panelist remarked that the \erb{} statement was so thorough it could serve as a prototype for future ethics statements.

\begin{table}[tb]
    \centering
    \begin{tabular}{>{\raggedright\arraybackslash}p{6.25cm}>{\raggedright\arraybackslash}p{1.5cm}}
         \textbf{Panelist action} & \textbf{Frequency} \\
         Identifying new risk & 26 \\
         Deepening the conversation & 22 \\
         No actions taken nor issues raised & 9 \\         
         Providing mitigation strategies & 6 \\
         Referring to specific previous work & 4 \\
         Identifying a potential collaborator & 2 \\
    \end{tabular}
    \caption{\erb{} panelists most commonly raised new risks that the researchers had not described in their statement, and deepened themes that the researchers had already raised.}
    \label{tab:actions}
\end{table}

As Table~\ref{tab:actions} illustrates, in addition to raising new risks and continuing the conversation, some panelists also provided specific mitigation strategies. In some cases, a panelist raised a new issue and outlined possible mitigation strategies for it; in others, the mitigation strategies pertained to ethical issues the researchers raised in their \erb{} statement but left insufficiently addressed. It was rarer for panelists to identify a potential collaborator for researchers to work with or refer the researchers to specific work on an issue. While some panelists had no feedback, at most this was only ever one of the two panelists for a project. This lends some confidence to the \erb{} process and its goals; PIs encountered some feedback on their projects, even if it passed review without additional actions needed.

A tension remains, however, between the risks identified in the \erb{} process and those that were not. It is possible that researchers and panelists alike are more attune to certain ethical issues than they are others, which may be a necessary constraint of the \erb{} pilot at a single university. However, while the most frequently cited issues revolved around similar themes between the researchers and the panelists, that is not to say that other ethical domains did not arise. For example, in one statement, researchers covered issues that ranged from consent and privacy to accountability and tool / user error. Similarly, in panelists' feedback for a project, issues covered publicness to representativeness.

\subsection{Researchers want even more scaffolding}
The most consistent constructive feedback from interviews was that researchers wanted the \erb{} not only to push them to broaden their ethical and societal lenses, but also to provide them with the scaffolding needed to make the appropriate considerations about their research. While the \erb{} statement prompt was kept brief due to a constraint from the funding program, several participants expressed concerns over the resulting vagueness. They requested more specificity, including additional examples of ethical violations in research or a workshop to help clarify the prompt.

\begin{quote}
    [The \erb{} didn’t] really help us figure out how to address these [ethical issues]... [They should] tell us how big the issues really are...the hard stuff is figuring out how important a particular ethical concern is. As researchers, we’re often left with trying to decide whether the positives outweigh the negatives in terms of use cases and ethics. What I found that the [\erb{}] didn't do was really help us in making those decisions about whether the positives outweigh the negatives or not. -~Researcher, medicine
\end{quote}

\begin{quote}
    It’d be nice if there [were] some foundational or bedrock things that were in [the statement prompt]. You know, one risk is [the statement] becomes template-y, which I think is a risk and a problem. But having to write another page when you're an academic is useful because it forces you to think these things through, which we've discussed, but it's just more burden. In my view the burden here is worth it but [if] there [were] some sort of help that would scaffold a researcher through rather than just, “okay, here's a blank page start from scratch.” -~Researcher, social science 
\end{quote}

\begin{quote}
    I wonder if there was a way of walking through [...] having “these are the kinds of considerations,” kind of a little more granularity might help another [\erb{}]. Perhaps a training module or links to examples, so that one could either go through the training or review examples, etc., as part of completing the [\erb{}]. -~Researcher, earth science
\end{quote}

\begin{quote}
    The prompt was more like, “what do you think is wrong?” And...I don't think anything is accomplished. You’re trying to figure out [how to] poke holes in your own story. It's very difficult. But … if there was more specificity … such as … [specific questions like] “does your thing discriminate on a protected characteristic?” -~Researcher, engineering
\end{quote}

Some interviewees recommended that the ESB utilize the statements from this deployment to compile a list of ethical issues that have arisen in projects. This list could serve as a resource for future applicants that participate in the ESB process. One respondent even suggested identifying some anonymized proposals as archetypes for “good statements,” especially those that contained specific action plans. In conjunction with more training and guidance from the ESB generally, adding more specificity to the \erb{} statement prompt was the main recommendation from four separate respondents.

The \erb{} panel feedback also can improve its scaffolding. It was useful for some: four interviewees, one of whom had iterated with the \erb{}, listed receiving feedback from the panelists as the most helpful aspect of the \erb{} process. The feedback provided insights about the ethical and societal implications of respondents’ work and substantively engaged with their \erb{} statements. Among those four respondents, one particularly appreciated the mutual respect shared with the panelists. They had received feedback from the \erb{} panelists inquiring about how they would share findings about sea-level research with nearby indigenous communities. Their response to the panelists indicated that this was not a practice in their field as proximity to ice sheets did not affect some communities’ sea level more than others. Panelists respected the respondent’s expertise about this and, consequently, did not push them to address the question further.

\begin{quote}
    I think it seemed like we were interacting with real people who understood us saying, “look, we understand what you're saying, that's not how sea level works” in a like “cool, okay [way].” That ability to communicate like a person was what made it okay. If, instead, it had been like, “look, we understand you say this, but this is a form. We have to fill [it] out and you're gonna have to come up with some sort of answer here.” And then we [would be] servicing a bureaucratic process as opposed to engaging with a responsive person. -~Researcher, earth science
\end{quote}

And for the others who iterated, 3 participants indicated that revising their \erb{} statement based on \erb{} feedback was the most helpful part of the process. In each case, respondents felt the iterative process gave them an opportunity to refine their thinking, clarify elements about their project they had omitted from the statement, and engage with feedback from \erb{} panelists. 

One the other hand, three interview respondents felt that the panel feedback needed more scaffolding. One indicated that the panel feedback did not help them prioritize the ethical issues nor confirm whether the issues were sufficiently mitigated. Another participant noted that, while the feedback didn’t engage much with their statement, it was not obvious the type of feedback to expect when the statement didn’t reveal any ethical issues needing mitigation. A third respondent felt that the panelists’ comments were similar to a paper review, commenting on details within the statement that could be revised but not pushing the respondent to engage with perspectives they hadn’t considered:
\begin{quote}
    I think if the [\erb{}] feedback was more about like the ethical implications of the [application package] as a whole, I think it would be more useful than if it was a feedback on the [\erb{} statement] narrowly, because I think the value of having this … very distinguished committee reading the [\erb{} statement] is that they have perspectives that I haven't thought about. So I want to hear those right, much more so than what details I missed in the things that I thought about. -~Researcher, engineering
\end{quote}

\begin{quote}
    I don't know exactly how the process in total worked, but maybe some others had actually a back and forth [with] more things to discuss. And I think that would be good. I would certainly appreciate more feedback ... if I had been thinking about it maybe not in the optimal way … I didn't have an expert for us. And I think that would be helpful, especially because I'm not an expert. -~Researcher, engineering
\end{quote}

\begin{quote}
    I also thought that the feedback didn't engage with the statement … Mostly the feedback that was provided was, “have you thought of X? Have you thought Y? Have you thought of Z?” and in our case, in particular, we had thought of most of those things and ... there was no even real strong sense of how big an issue the board felt it was, which is a huge issue for me because you know, the way those were written, sometimes, there was … very, very strong concern. -~Researcher, engineering
\end{quote}

Among those who iterated with the \erb{}, one respondent was frustrated that they didn’t see the difference between the IRB and the \erb{} process as they experienced it, which made the process feel like a bureaucratic requirement.

\begin{quote}
    We submitted a response to their original response. And then they responded, then we responded to it again. And it was in our final response that I kind of felt like it was a little bit like busy work. I just felt like … the questions weren't going to change our research design in any way. They just wanted us to have a response to this manner that they raised and that felt very IRB to me. I guess maybe I don't really understand what the difference is supposed to be between [\erb{}] and the IRB. Because I think the IRB would say they're trying to do the same thing as an [\erb{}], to get you to think through, you know, ethical issues and possibly adjust your project design, as ... issues arise. So I guess it, to me, it kind of felt like some of this was, you know, a little … redundant.
    
    And I think [the \erb{}] should also really think through their relationship … how they're different from the IRB because I feel like this was pretty similar to the IRB process, just at a different stage … And it increases the feeling that thinking about ethics as busy work when I wish it was … not that way. -~Researcher, engineering
    
\end{quote}

One possible remedy to this situation---at more time cost to the panel---would be to engage in conversations rather than written feedback.

\subsection{Institutional reflection: power and consistency}
\begin{figure}[tb]
    \centering
    \includegraphics[width=\columnwidth]{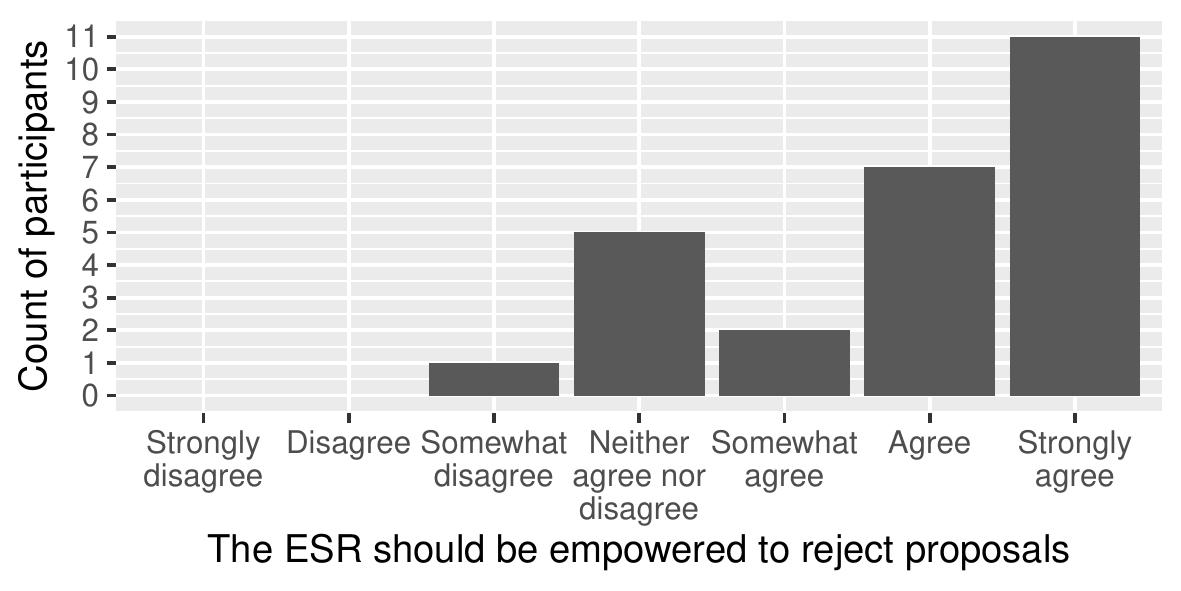}
    \caption{The general trend from researchers was that the \erb{} should be empowered to reject proposals if necessary.}
    \Description[Histogram of whether the ESR should be empowered to reject proposals]{A histogram, with X axis labeled ``The ESR should be empowered to reject proposals''. Eleven responses ``strongly agree', seven responses ``agree'', two responses ``somewhat agree'', five responses ``neither agree nor disagree'', and one response ``somewhat disagree''.}
    \label{fig:rejection}
\end{figure}

A potentially contentious issue is whether a panel like the \erb{} should be empowered to deny funding to a project. There was generally a consensus via the survey that this was desirable, and no moderate or strong disagreement (Figure~\ref{fig:rejection}). The story was more complex within the interviews. Amongst interviewees, although 11 agreed to varying degrees that the \erb{} should be empowered to reject an especially ethically problematic proposal, 5 of those participants strongly encouraged the \erb{} to prioritize the iterative process over a brute enforcement mechanism. They preferred that all participants were invited to iterate their initial statements with the \erb{} panelists. Only after some predetermined and documented point should a participant face rejection if they do not demonstrate a willingness to engage with panelists’ recommendations and feedback. 

However, other respondents, primarily those in Engineering, were more hesitant to the idea of vesting enforcement mechanisms within the \erb{}. One raised questions about the role of the \erb{} if a project is deemed harmful down the line.
\begin{quote}
    I think one thing to think about is … whether you know if you're certified by the [\erb{}] is your project ethically sound … The IRB has a similar question and … I assume that if they pass too many experiments that later turn out to be sort of dubious or harmful that board will be in real trouble. It doesn't feel like the [\erb{}] currently would have those consequences [in] place. And without those consequences in place, without the [\erb{}] really certifying and taking arrows, if something blows up … it’s not really clear to me how much authority they should have in deciding whether a project is fundable or not. -~Researcher, medicine
\end{quote}

Others were concerned that societal good would be factored into funding decisions or fears over potential harms could prevent researchers from examining important questions.
\begin{quote}
    But I worry about the world in which the [\erb{}] is sort of empowered to reject proposals and then now [the application review] becomes sort of a weighted score [of] scientific merit plus like societal good. -~Researcher, engineering 
\end{quote}

\begin{quote}
    I think that these [potentially problematic research projects] are things that are going to happen. And so I would not like to see just, you know, trying to prevent it from happening. I think it's more important … to investigate those issues than it is to try to prevent people from executing them.  I guess there could be … situations where the research itself is problematic … but I think just because there are ethical concerns with the outcome of the research, [it probably makes it] even more important to look into that research. -~Researcher, engineering
\end{quote}

\begin{quote}
    If there is such a board that has the ability to stop the research, then the bar for what it means to stop it, and why should be high, or if you can immediately show there are people being harmed by this. There are protected characteristics that are being harmed by this right, then by all means, that needs to be stopped. But if there's concern about what something could be, you have to have faith that the questions we ask will ultimately lead to a positive outcome.  -~Researcher, engineering
\end{quote}

When considering \erb{} authority to reject a proposal over its potential harms, 4 respondents mentioned weighing those costs against the potential benefits of the research, similar to the framework used by the IRB. All of those respondents expected this cost-benefit calculation to be the \erb{}’s responsibility.
\begin{quote}
    Often as investigators, we’re not qualified to make that decision. And I think that having a body of experts and power to make the decision, it's kind of a safeguard … for us, for the public. -~Researcher, earth science 
\end{quote}

\section{Discussion}
Overall, PIs identified the \erb{} experience as positive in providing scaffolding for ethical reflection, and desired even more scaffolding. In this section, we reflect on our experiences and learning over the past year.

\subsection{Scaling, growth, and replicability}
We are currently partnering with other funding organizations on our campus to grow the footprint of the \erb{} beyond AI research. Other areas of research, including research in elections, bioengineering, data science, and sustainability have expressed interest. Growing the size and scope of the \erb{}, e.g., from 40 proposals to 100 or 400, will require changes in process so as not to overwhelm the faculty panel. For example, it may be possible for a staff panel, or qualified postdoctoral scholars or graduate students, to perform a first triage round of filtering. Based on our experiences this year, a triage round could likely reduce the number of proposals reviewed by faculty by about half, as many grants are not of substantial ethical or societal risk. These proposals could still receive written feedback on any areas that the staff identified, and an opportunity to engage in a conversation with faculty if the researchers desire a consult. 

More ambitously, we hope to spread the practice of an \erb{} to other universities and organizations. We hope that other universities could adopt a similar process. Toward this end, we will be sharing materials that include submission requirements, support scaffolding for researchers, review procedures and rubrics for panelists, and example case studies. Ultimately, this can only succeed if we are able to make the benefit substantial and the effort required to start this process by faculty at other institutions manageable. These procedures are still in early days: we believe it would be beneficial for the community to convene workshops and discussions to share experiences, challenges, and best practices.

We have begun fielding interest from industry, who wish to create similar structures. Some technology companies already have committees that explore similar roles, such as Microsoft's Aether. Why would a company want to include this? First, there are obvious normative reasons to include an \erb{}. However, approaches such as the \erb{} also manage risk: existential risk, reputational risk, and the risk of loss of trust. They also position the organization such that, no matter any upcoming regulation, the company will already be ahead of it and thereby gain a competitive advantage.

As the process grows, it will be important to understand its own internal consistency. If the same proposal is submitted to different subsets of the panel, will it receive similar feedback each time? Evidence from the pilot suggests that the \emph{level} of concern will likely be similar, but a study of inter-rater reliability on specific categories of concern would help.

Ultimately, as processes such as the \erb{} prove themselves out, the question arises of whether they can be adopted by computing professional societies such as the Association for Computing Machinery (ACM). The ACM already has a code of ethics, but few processes of ethics.

\subsection{Process improvements}
Clearly, one of the tightrope walks with an \erb{} process is formalization vs.\ flexibility. PIs desired formalization: what issues they should worry about, and how to handle them. However, over-formalization can become associated with heavyweight bureacracy, and restrict the flexibility needed to consider the ethical and social challenges of each individual project, especially given that the \erb{} is not currently beholden to regulatory requirements. Pragmatically, many PIs want more materials, but our sense so far was that many of these same PIs did not spend time perusing the materials that we had already assembled. To strike a path forward, we have drawn on outcomes from this first year. In particular, our instructions for the second year of the program now include a list of the most common issues raised by panelists as well as example principles for mitigation and resulting research designs. As future work, we feel that it will be important to articulate a broader set of guidelines for panelists to use in calibrating their feedback.

An intriguing alternative would be to structure the process less like a review and more like a conversation. This might help orient the feedback less toward negatives. For example, instead of asking for a written response from every PI whose proposal is not immediately passed, the \erb{} could set up a brief meeting between \erb{} representatives and the panelists. We hope that this might create a less antagonistic assumption, and also differentiate more meaningfully from IRB structures.

As the \erb{} grows and matures, it will be increasingly important to ask how heavyhanded the program will be. The program relies on the faculty involved striking a balance between feedback and pressure, especially since the \erb{} can delay access to funding or recommend against funding. At least one member of our panel was interested in vetoing several projects. Conversely, if the panel is too light a touch, the process can become subverted and without teeth, a process of ethicswashing~\cite{bietti2020ethics}. Researchers have largely accepted the need for an IRB review, though opinions are mixed on it. Our hope is the the \erb{} process can be seen as a benefit.

Can and how should the \erb{} transition from \textit{ex ante} review to ongoing feedback? One intriguing possibility is that many funders require annual reports on their grants. We are currently coordinating with the funding program to request a brief update on the project as it impacts the \erb{} statement that the team originally submitted. Has the project changed in ways that would benefit from additional conversation or review?

\subsection{Possible negative outcomes of an \erb{}}
It also seems prudent that an effort that's focused on identifying risks to society should also introspect on the risks that it itself presents. One concern is: what happens when the \erb{} misses something? For example, a protocol that the ESB approves for developing algorithms for healthcare triage might later be found to accidentally encode racial bias~\cite{obermeyer2019dissecting}. The \erb{} ideally would catch this and ask the researchers to commit to checking; however, the \erb{} might not catch it, and the researchers might not think to check across other axes and get called out later as well. This would inevitably cause the public to criticize both the project but also the ESB. Ultimately, the ESB aims to make more unforeseen risks become foreseen, but it will not be an oracle. In these situations, the ESB might make a short statement acknowledging the issue, and reflect on how to do better next time.

Another concern is: is academic freedom being threatened? We argue not, because researchers are still free to pursue their research through other funding sources. However, we should imagine a potential future where this becomes more widespread, and consider ourselves with the outcomes there. No paragraph in a Discussion section can fully address academic freedom debates, and the nature of ethics is such that new cases require new consideration, so we cannot make unilateral promises. But, we do think that to mitigate these concerns, it will be important for the \erb{} to publish a clear set of principles and guidelines that it uses in discussion. In our case, this has been about the translation of harms into principles into mitigation as a requirement for participation in the \erb{} process. However, drawing the line on what mitigation is sufficient remains an open question.

A third concern: who gets to decide who is the arbiter of what is ethical, and why? If a project struggles to get approval from the \erb{}, it will produce tension and possibly accusations of stonewalling. The best approach that we can advocate for here is to ensure that the panel consists of a cross-section of respected and relevant faculty from across campus: so, while the decision may be frustrating, the people who make it are reputable and considerate. This will be required in order to lend legitimacy to the effort.
\section{Conclusion}
In this paper, we present an approach to supporting early-stage feedback on societal and ethical dimensions of AI research. The process, which we termed Ethics and Society Board (\erb{}), inserts itself into the grant funding process as a lever for reflection. By drawing on a cross-campus interdisciplinary panel, researchers have a chance to engage in this reflection early on in the lifecycle of their projects. Moving forward, we are fielding interest in expanding this program beyond AI research. Other areas, such as public interest technology, sustainability, bioengineering and data science face similar questions.

\section{Acknowledgments}
This work was supported by the Public Interest Technology University Network; Stanford's Ethics, Science, and Technology Hub; and Stanford's Institute for Human-Centered Artificial Intelligence. Thank you to Ashlyn Jaeger, James Landay, Fei-Fei Li, John Etchemendy, Deep Ganguli, and Vanessa Parli for their support. Thank you to the Stanford Institute for Human-Centered Artificial Intelligence for their enthusiasm and collaboration. Thank you to the faculty panelists on the \erb{} for their time, insight, and energy. Thank you to the researchers who engaged with the \erb{} for their effort and feedback. Thank you to Adam Bailey at Stanford's Institutional Review Board for his advice, and to Mary Gray at Microsoft Research for her leadership and feedback.

\appendix
\section{Appendix: Interview Protocol}
Thank you for agreeing to this conversation. Just to remind you of our goal, we’re trying to evaluate the \erb{} process and get your thoughts on how the process went. Nothing you say today will be quoted with your name. We’ll be submitting a report to the HAI executive committee and publishing our results in an academic journal, but no confidential names will be mentioned. Do you have any questions about our pledge of confidentiality?

First, we would like to ask you some questions about your experience considering the ethical implications of your work and your ethics training.
\begin{enumerate}
    \item How much do you typically engage with ethical issues in the research you do?
    \item How much do researchers in your field take into account the ethical implications of their research? 
    \item Do you believe that most faculty are sufficiently trained to consider the ethical issues with their research? 
    \item Graduate students?
\end{enumerate}

Next, we would like to ask you about your experience with the \erb{} process. 
\begin{enumerate}
    \item Through the \erb{} process, did you become aware of ethical issues you hadn't previously considered? 
    \item Why (why not) was the \erb{} able to make you aware of these issues? / Which aspects of the \erb{} process alerted you to these issues? 
    \item How helpful to you was the identification of these issues? 
    \item Did it lead to concrete changes in your project design?
    \item What would you change about the \erb{} process? 
\end{enumerate}
 
Now we would like to follow-up on a few questions we asked in the survey you completed after the \erb{} review process. 
\begin{enumerate}
    \item We had asked you, “Which aspect of the \erb{} process did you find least helpful?”, you responded that [response] was the least helpful.  Why was that the least helpful aspect of the process for you? 
    \item We had asked you, “Which aspect of the \erb{} process did you find most helpful?”, you responded that [response] was the most helpful.  Why was that the most helpful aspect of the process for you? 
    \item Research projects occasionally garner public criticism (e.g., in news media or social media) for ethical missteps. Examples in the last few years at Stanford included election experiments, engineering innovations that encourage worker displacement, and dataset bias. In our survey, we asked you, “How strongly do you agree with the claim: \erb{} review helped mitigate the possibility of public criticism of my project." You responded that [response]. Why do you [response]?
\end{enumerate}

Finally, we would like to get your thoughts on the authority the \erb{} should have and your willingness to engage in ethical reviews in the future. 
\begin{enumerate}
    \item The \erb{}’s main purpose is to work with researchers to reflect on and design ethics into their research. How would you say your experience aligns with the intended purpose of the \erb{}? [Question added to interview protocols used after December 24, 2020]
    \item The \erb{}’s main purpose is to work with researchers to reflect on and design ethics into their research. However, the \erb{} is also called on to make decisions about when to decline to recommend a project for funding due to ethical concerns. In our survey, we asked you, "How strongly do you agree with the claim: the \erb{} should be empowered to reject an especially ethically problematic proposal." You responded that [response]. Why do you [response]?
    \item Under what conditions should the \erb{} not recommend, if ever, a project for funding?
    \item Are there any circumstances where the \erb{} should recommend a project for funding even though considerable potential harms have been identified?
    \item Under what conditions do you think projects should be reviewed by an ethics review panel? 
    \item Should the \erb{} make summaries of their recommendations publicly available? 
    \item Under what conditions would you be willing to submit a future research project to the \erb{} for review? 
    \item Is there anything else you would like to share about your experience with the \erb{}? 
\end{enumerate}

\bibliography{references}

\begin{thebibliography}{10}

\bibitem{abuhamad2020like}
{\sc Abuhamad, G., and Rheault, C.}
\newblock Like a researcher stating broader impact for the very first time.
\newblock {\em arXiv preprint arXiv:2011.13032\/} (2020).

\bibitem{agarwal2018reductions}
{\sc Agarwal, A., Beygelzimer, A., Dud{\'\i}k, M., Langford, J., and Wallach,
  H.}
\newblock A reductions approach to fair classification.
\newblock {\em arXiv preprint arXiv:1803.02453\/} (2018).

\bibitem{alkhatib2021utopia}
{\sc Alkhatib, A.}
\newblock To live in their utopia: Why algorithmic systems create absurd
  outcomes.
\newblock In {\em Proceedings of the 2021 CHI Conference on Human Factors in
  Computing Systems\/} (New York, NY, USA, 2021), CHI '21, Association for
  Computing Machinery.

\bibitem{alkhatib2019street}
{\sc Alkhatib, A., and Bernstein, M.}
\newblock Street-level algorithms: A theory at the gaps between policy and
  decisions.
\newblock In {\em Proceedings of the 2019 CHI Conference on Human Factors in
  Computing Systems\/} (2019), pp.~1--13.

\bibitem{bamberger2015privacy}
{\sc Bamberger, K.~A., and Mulligan, D.~K.}
\newblock {\em Privacy on the ground: driving corporate behavior in the United
  States and Europe}.
\newblock MIT Press, 2015.

\bibitem{bao2021s}
{\sc Bao, M., Zhou, A., Zottola, S., Brubach, B., Desmarais, S., Horowitz, A.,
  Lum, K., and Venkatasubramanian, S.}
\newblock It's compaslicated: The messy relationship between rai datasets and
  algorithmic fairness benchmarks.
\newblock {\em arXiv preprint arXiv:2106.05498\/} (2021).

\bibitem{barocas2014big}
{\sc Barocas, S., and Nissenbaum, H.}
\newblock Big data’s end run around anonymity and consent.
\newblock {\em Privacy, big data, and the public good: Frameworks for
  engagement 1\/} (2014), 44--75.

\bibitem{barocas2016big}
{\sc Barocas, S., and Selbst, A.~D.}
\newblock Big data's disparate impact.
\newblock {\em Calif. L. Rev. 104\/} (2016), 671.

\bibitem{belfield2020activism}
{\sc Belfield, H.}
\newblock Activism by the ai community: analysing recent achievements and
  future prospects.
\newblock In {\em Proceedings of the AAAI/ACM Conference on AI, Ethics, and
  Society\/} (2020), pp.~15--21.

\bibitem{benjamin2019race}
{\sc Benjamin, R.}
\newblock Race after technology: Abolitionist tools for the new jim code.
\newblock {\em Social Forces\/} (2019).

\bibitem{bietti2020ethics}
{\sc Bietti, E.}
\newblock From ethics washing to ethics bashing: a view on tech ethics from
  within moral philosophy.
\newblock In {\em Proceedings of the 2020 Conference on Fairness,
  Accountability, and Transparency\/} (2020), pp.~210--219.

\bibitem{binns2018s}
{\sc Binns, R., Van~Kleek, M., Veale, M., Lyngs, U., Zhao, J., and Shadbolt,
  N.}
\newblock `it's reducing a human being to a percentage': Perceptions of justice
  in algorithmic decisions.
\newblock In {\em Proceedings of the 2018 Chi conference on human factors in
  computing systems\/} (2018), pp.~1--14.

\bibitem{bliss2020agenda}
{\sc Bliss, N., Bradley, E., Garland, J., Menczer, F., Ruston, S.~W., Starbird,
  K., and Wiggins, C.}
\newblock An agenda for disinformation research.
\newblock {\em arXiv preprint arXiv:2012.08572\/} (2020).

\bibitem{blodgett2020language}
{\sc Blodgett, S.~L., Barocas, S., Daum{\'e}~III, H., and Wallach, H.}
\newblock Language (technology) is power: A critical survey of ``bias'' in nlp.
\newblock {\em arXiv preprint arXiv:2005.14050\/} (2020).

\bibitem{boyarskaya2020overcoming}
{\sc Boyarskaya, M., Olteanu, A., and Crawford, K.}
\newblock Overcoming failures of imagination in ai infused system development
  and deployment, 2020.

\bibitem{buolamwini2018gender}
{\sc Buolamwini, J., and Gebru, T.}
\newblock Gender shades: Intersectional accuracy disparities in commercial
  gender classification.
\newblock In {\em Conference on fairness, accountability and transparency\/}
  (2018), pp.~77--91.

\bibitem{burton2018}
{\sc Burton, E., Goldsmith, J., and Mattei, N.}
\newblock How to teach computer ethics through science fiction.
\newblock {\em Commun. ACM 61}, 8 (July 2018), 54–64.

\bibitem{calvo2018ai}
{\sc Calvo, R.~A., and Peters, D.}
\newblock Ai surveillance studies need ethics review.
\newblock {\em Nature 557}, 7706 (2018), 31--32.

\bibitem{canosa2008}
{\sc Canosa, R.~L., and Lucas, J.~M.}
\newblock Mock trials and role-playing in computer ethics courses.
\newblock {\em SIGCSE Bull. 40}, 1 (Mar. 2008), 148–152.

\bibitem{char2018implementing}
{\sc Char, D.~S., Shah, N.~H., and Magnus, D.}
\newblock Implementing machine learning in health care—addressing ethical
  challenges.
\newblock {\em The New England journal of medicine 378}, 11 (2018), 981.

\bibitem{chouldechova2017fair}
{\sc Chouldechova, A.}
\newblock Fair prediction with disparate impact: A study of bias in recidivism
  prediction instruments.
\newblock {\em Big data 5}, 2 (2017), 153--163.

\bibitem{christin2017algorithms}
{\sc Christin, A.}
\newblock Algorithms in practice: Comparing web journalism and criminal
  justice.
\newblock {\em Big Data \& Society 4}, 2 (2017), 2053951717718855.

\bibitem{corbett2017algorithmic}
{\sc Corbett-Davies, S., Pierson, E., Feller, A., Goel, S., and Huq, A.}
\newblock Algorithmic decision making and the cost of fairness.
\newblock In {\em Proceedings of the 23rd acm sigkdd international conference
  on knowledge discovery and data mining\/} (2017), pp.~797--806.

\bibitem{costanza2018design}
{\sc Costanza-Chock, S.}
\newblock Design justice: towards an intersectional feminist framework for
  design theory and practice.
\newblock {\em Proceedings of the Design Research Society\/} (2018).

\bibitem{belmont1979}
{\sc {Department of Health, Education, and Welfare}}.
\newblock {The Belmont Report}: Ethical principles and guidelines for the
  protection of human subjects of research, 1979.

\bibitem{diaz2018addressing}
{\sc D{\'\i}az, M., Johnson, I., Lazar, A., Piper, A.~M., and Gergle, D.}
\newblock Addressing age-related bias in sentiment analysis.
\newblock In {\em Proceedings of the 2018 CHI Conference on Human Factors in
  Computing Systems\/} (2018), pp.~1--14.

\bibitem{dwork2012fairness}
{\sc Dwork, C., Hardt, M., Pitassi, T., Reingold, O., and Zemel, R.}
\newblock Fairness through awareness.
\newblock In {\em Proceedings of the 3rd innovations in theoretical computer
  science conference\/} (2012), pp.~214--226.

\bibitem{eubanks2018automating}
{\sc Eubanks, V.}
\newblock {\em Automating inequality: How high-tech tools profile, police, and
  punish the poor}.
\newblock St. Martin's Press, 2018.

\bibitem{fazelpour2020algorithmic}
{\sc Fazelpour, S., and Lipton, Z.~C.}
\newblock Algorithmic fairness from a non-ideal perspective.
\newblock In {\em Proceedings of the AAAI/ACM Conference on AI, Ethics, and
  Society\/} (2020), pp.~57--63.

\bibitem{fiesler2020}
{\sc Fiesler, C., Garrett, N., and Beard, N.}
\newblock What do we teach when we teach tech ethics? a syllabi analysis.
\newblock In {\em Proceedings of the 51st ACM Technical Symposium on Computer
  Science Education\/} (New York, NY, USA, 2020), SIGCSE '20, Association for
  Computing Machinery, p.~289–295.

\bibitem{johnson2021communitypanel}
{\sc G~Johnson, I., and Crivellaro, C.}
\newblock Opening research commissioning to civic participation: Creating a
  community panel to review the social impact of hci research proposals.
\newblock In {\em Proceedings of the 2021 CHI Conference on Human Factors in
  Computing Systems\/} (New York, NY, USA, 2021), CHI '21, Association for
  Computing Machinery.

\bibitem{garg2018word}
{\sc Garg, N., Schiebinger, L., Jurafsky, D., and Zou, J.}
\newblock Word embeddings quantify 100 years of gender and ethnic stereotypes.
\newblock {\em Proceedings of the National Academy of Sciences 115}, 16 (2018),
  E3635--E3644.

\bibitem{gebru2018datasheets}
{\sc Gebru, T., Morgenstern, J., Vecchione, B., Vaughan, J.~W., Wallach, H.,
  Daum{\'e}~III, H., and Crawford, K.}
\newblock Datasheets for datasets.
\newblock {\em arXiv preprint arXiv:1803.09010\/} (2018).

\bibitem{gibney2018ethics}
{\sc Gibney, E.}
\newblock The ethics of computer science: this researcher has a controversial
  proposal.
\newblock {\em Nature 26\/} (2018).

\bibitem{gray2019ghost}
{\sc Gray, M.~L., and Suri, S.}
\newblock {\em Ghost work: how to stop Silicon Valley from building a new
  global underclass}.
\newblock Eamon Dolan Books, 2019.

\bibitem{grosz2019}
{\sc Grosz, B.~J., Grant, D.~G., Vredenburgh, K., Behrends, J., Hu, L.,
  Simmons, A., and Waldo, J.}
\newblock Embedded ethics: Integrating ethics across cs education.
\newblock {\em Commun. ACM 62}, 8 (July 2019), 54–61.

\bibitem{hardt2016equality}
{\sc Hardt, M., Price, E., and Srebro, N.}
\newblock Equality of opportunity in supervised learning.
\newblock In {\em Advances in neural information processing systems\/} (2016),
  pp.~3315--3323.

\bibitem{hartmann2020next}
{\sc Hartmann, K., and Giles, K.}
\newblock The next generation of cyber-enabled information warfare.
\newblock In {\em 2020 12th International Conference on Cyber Conflict
  (CyCon)\/} (2020), vol.~1300, IEEE, pp.~233--250.

\bibitem{holstein2019improving}
{\sc Holstein, K., Wortman~Vaughan, J., Daum{\'e}~III, H., Dudik, M., and
  Wallach, H.}
\newblock Improving fairness in machine learning systems: What do industry
  practitioners need?
\newblock In {\em Proceedings of the 2019 CHI Conference on Human Factors in
  Computing Systems\/} (2019), pp.~1--16.

\bibitem{hudson2004go}
{\sc Hudson, J.~M., and Bruckman, A.}
\newblock “go away”: participant objections to being studied and the ethics
  of chatroom research.
\newblock {\em The Information Society 20}, 2 (2004), 127--139.

\bibitem{jobin2019global}
{\sc Jobin, A., Ienca, M., and Vayena, E.}
\newblock The global landscape of ai ethics guidelines.
\newblock {\em Nature Machine Intelligence 1}, 9 (2019), 389--399.

\bibitem{jordan2019designing}
{\sc Jordan, S.~R.}
\newblock Designing an artificial intelligence research review committee, 2019.

\bibitem{kaminski2020multi}
{\sc Kaminski, M.~E., and Malgieri, G.}
\newblock Multi-layered explanations from algorithmic impact assessments in the
  gdpr.
\newblock In {\em Proceedings of the 2020 Conference on Fairness,
  Accountability, and Transparency\/} (2020), pp.~68--79.

\bibitem{kay2015unequal}
{\sc Kay, M., Matuszek, C., and Munson, S.~A.}
\newblock Unequal representation and gender stereotypes in image search results
  for occupations.
\newblock In {\em Proceedings of the 33rd Annual ACM Conference on Human
  Factors in Computing Systems\/} (2015), pp.~3819--3828.

\bibitem{ko2020}
{\sc Ko, A.~J., Oleson, A., Ryan, N., Register, Y., Xie, B., Tari, M.,
  Davidson, M., Druga, S., and Loksa, D.}
\newblock It is time for more critical cs education.
\newblock {\em Commun. ACM 63}, 11 (Oct. 2020), 31–33.

\bibitem{krafft2020defining}
{\sc Krafft, P., Young, M., Katell, M., Huang, K., and Bugingo, G.}
\newblock Defining ai in policy versus practice.
\newblock In {\em Proceedings of the AAAI/ACM Conference on AI, Ethics, and
  Society\/} (2020), pp.~72--78.

\bibitem{krawiec2003cosmetic}
{\sc Krawiec, K.~D.}
\newblock Cosmetic compliance and the failure of negotiated governance.
\newblock {\em Wash. ULQ 81\/} (2003), 487.

\bibitem{kulshrestha2017quantifying}
{\sc Kulshrestha, J., Eslami, M., Messias, J., Zafar, M.~B., Ghosh, S.,
  Gummadi, K.~P., and Karahalios, K.}
\newblock Quantifying search bias: Investigating sources of bias for political
  searches in social media.
\newblock In {\em Proceedings of the 2017 ACM Conference on Computer Supported
  Cooperative Work and Social Computing\/} (2017), pp.~417--432.

\bibitem{kusner2017counterfactual}
{\sc Kusner, M.~J., Loftus, J., Russell, C., and Silva, R.}
\newblock Counterfactual fairness.
\newblock In {\em Advances in neural information processing systems\/} (2017),
  pp.~4066--4076.

\bibitem{lee2018understanding}
{\sc Lee, M.~K.}
\newblock Understanding perception of algorithmic decisions: Fairness, trust,
  and emotion in response to algorithmic management.
\newblock {\em Big Data \& Society 5}, 1 (2018), 2053951718756684.

\bibitem{lee2017algorithmic}
{\sc Lee, M.~K., and Baykal, S.}
\newblock Algorithmic mediation in group decisions: Fairness perceptions of
  algorithmically mediated vs. discussion-based social division.
\newblock In {\em Proceedings of the 2017 ACM Conference on Computer Supported
  Cooperative Work and Social Computing\/} (2017), pp.~1035--1048.

\bibitem{madaio2020co}
{\sc Madaio, M.~A., Stark, L., Wortman~Vaughan, J., and Wallach, H.}
\newblock Co-designing checklists to understand organizational challenges and
  opportunities around fairness in ai.
\newblock In {\em Proceedings of the 2020 CHI Conference on Human Factors in
  Computing Systems\/} (2020), pp.~1--14.

\bibitem{mcinnis2016taking}
{\sc McInnis, B., Cosley, D., Nam, C., and Leshed, G.}
\newblock Taking a hit: Designing around rejection, mistrust, risk, and
  workers' experiences in amazon mechanical turk.
\newblock In {\em Proceedings of the 2016 CHI conference on human factors in
  computing systems\/} (2016), pp.~2271--2282.

\bibitem{merton1936unanticipated}
{\sc Merton, R.~K.}
\newblock The unanticipated consequences of purposive social action.
\newblock {\em American sociological review 1}, 6 (1936), 894--904.

\bibitem{metcalf2016human}
{\sc Metcalf, J., and Crawford, K.}
\newblock Where are human subjects in big data research? the emerging ethics
  divide.
\newblock {\em Big Data \& Society 3}, 1 (2016), 2053951716650211.

\bibitem{metcalf2019owning}
{\sc Metcalf, J., Moss, E., et~al.}
\newblock Owning ethics: Corporate logics, silicon valley, and the
  institutionalization of ethics.
\newblock {\em Social Research: An International Quarterly 86}, 2 (2019),
  449--476.

\bibitem{mitchell2020diversity}
{\sc Mitchell, M., Baker, D., Moorosi, N., Denton, E., Hutchinson, B., Hanna,
  A., Gebru, T., and Morgenstern, J.}
\newblock Diversity and inclusion metrics in subset selection.
\newblock In {\em Proceedings of the AAAI/ACM Conference on AI, Ethics, and
  Society\/} (2020), pp.~117--123.

\bibitem{mitchell2019model}
{\sc Mitchell, M., Wu, S., Zaldivar, A., Barnes, P., Vasserman, L., Hutchinson,
  B., Spitzer, E., Raji, I.~D., and Gebru, T.}
\newblock Model cards for model reporting.
\newblock In {\em Proceedings of the conference on fairness, accountability,
  and transparency\/} (2019), pp.~220--229.

\bibitem{mittelstadt2019ai}
{\sc Mittelstadt, B.}
\newblock Ai ethics--too principled to fail.
\newblock {\em arXiv preprint arXiv:1906.06668\/} (2019).

\bibitem{nanayakkara2021unpacking}
{\sc Nanayakkara, P., Hullman, J., and Diakopoulos, N.}
\newblock Unpacking the expressed consequences of ai research in broader impact
  statements.
\newblock {\em arXiv preprint arXiv:2105.04760\/} (2021).

\bibitem{noble2018algorithms}
{\sc Noble, S.~U.}
\newblock {\em Algorithms of oppression: How search engines reinforce racism}.
\newblock nyu Press, 2018.

\bibitem{obermeyer2019dissecting}
{\sc Obermeyer, Z., Powers, B., Vogeli, C., and Mullainathan, S.}
\newblock Dissecting racial bias in an algorithm used to manage the health of
  populations.
\newblock {\em Science 366}, 6464 (2019), 447--453.

\bibitem{o2016weapons}
{\sc O'neil, C.}
\newblock {\em Weapons of math destruction: How big data increases inequality
  and threatens democracy}.
\newblock Broadway Books, 2016.

\bibitem{orlikowski1992duality}
{\sc Orlikowski, W.~J.}
\newblock The duality of technology: Rethinking the concept of technology in
  organizations.
\newblock {\em Organization science 3}, 3 (1992), 398--427.

\bibitem{raji2020closing}
{\sc Raji, I.~D., Smart, A., White, R.~N., Mitchell, M., Gebru, T., Hutchinson,
  B., Smith-Loud, J., Theron, D., and Barnes, P.}
\newblock Closing the ai accountability gap: defining an end-to-end framework
  for internal algorithmic auditing.
\newblock In {\em Proceedings of the 2020 Conference on Fairness,
  Accountability, and Transparency\/} (2020), pp.~33--44.

\bibitem{rakova2020responsible}
{\sc Rakova, B., Yang, J., Cramer, H., and Chowdhury, R.}
\newblock Where responsible ai meets reality: Practitioner perspectives on
  enablers for shifting organizational practices.
\newblock {\em arXiv preprint arXiv:2006.12358\/} (2020).

\bibitem{register2020}
{\sc Register, Y., and Ko, A.~J.}
\newblock Learning machine learning with personal data helps stakeholders
  ground advocacy arguments in model mechanics.
\newblock In {\em Proceedings of the 2020 ACM Conference on International
  Computing Education Research\/} (New York, NY, USA, 2020), ICER '20,
  Association for Computing Machinery, p.~67–78.

\bibitem{reisman2018algorithmic}
{\sc Reisman, D., Schultz, J., Crawford, K., and Whittaker, M.}
\newblock Algorithmic impact assessments: A practical framework for public
  agency accountability.
\newblock {\em AI Now Institute\/} (2018), 1--22.

\bibitem{robertson2021modeling}
{\sc Robertson, S., Nguyen, T., and Salehi, N.}
\newblock Modeling assumptions clash with the real world: Transparency, equity,
  and community challenges for student assignment algorithms.
\newblock In {\em Proceedings of the 2021 CHI Conference on Human Factors in
  Computing Systems\/} (2021), pp.~1--14.

\bibitem{saltz2019}
{\sc Saltz, J., Skirpan, M., Fiesler, C., Gorelick, M., Yeh, T., Heckman, R.,
  Dewar, N., and Beard, N.}
\newblock Integrating ethics within machine learning courses.
\newblock {\em ACM Trans. Comput. Educ. 19}, 4 (Aug. 2019).

\bibitem{sandvig2014auditing}
{\sc Sandvig, C., Hamilton, K., Karahalios, K., and Langbort, C.}
\newblock Auditing algorithms: Research methods for detecting discrimination on
  internet platforms.
\newblock {\em Data and discrimination: converting critical concerns into
  productive inquiry 22\/} (2014).

\bibitem{selbst2019fairness}
{\sc Selbst, A.~D., boyd, d., Friedler, S.~A., Venkatasubramanian, S., and
  Vertesi, J.}
\newblock Fairness and abstraction in sociotechnical systems.
\newblock In {\em Proceedings of the Conference on Fairness, Accountability,
  and Transparency\/} (2019), pp.~59--68.

\bibitem{shneiderman2020human}
{\sc Shneiderman, B.}
\newblock Human-centered artificial intelligence: Reliable, safe \&
  trustworthy.
\newblock {\em International Journal of Human--Computer Interaction 36}, 6
  (2020), 495--504.

\bibitem{skirpan2018}
{\sc Skirpan, M., Beard, N., Bhaduri, S., Fiesler, C., and Yeh, T.}
\newblock Ethics education in context: A case study of novel ethics activities
  for the cs classroom.
\newblock In {\em Proceedings of the 49th ACM Technical Symposium on Computer
  Science Education\/} (New York, NY, USA, 2018), SIGCSE '18, Association for
  Computing Machinery, p.~940–945.

\bibitem{ethicssurvey2016}
{\sc Stahl, B.~C., Timmermans, J., and Mittelstadt, B.~D.}
\newblock The ethics of computing: A survey of the computing-oriented
  literature.
\newblock {\em ACM Comput. Surv. 48}, 4 (Feb. 2016).

\bibitem{toyama2015geek}
{\sc Toyama, K.}
\newblock {\em Geek heresy: Rescuing social change from the cult of
  technology}.
\newblock PublicAffairs, 2015.

\bibitem{trevino1999managing}
{\sc Trevino, L.~K., Weaver, G.~R., Gibson, D.~G., and Toffler, B.~L.}
\newblock Managing ethics and legal compliance: What works and what hurts.
\newblock {\em California management review 41}, 2 (1999), 131--151.

\bibitem{commonrule}
{\sc {United States Department of Health and Human Services}}.
\newblock Common rule.
\newblock {\em Code of Federal Regulations Title 45\/} (2018), §46.111.

\bibitem{veale2018fairness}
{\sc Veale, M., Van~Kleek, M., and Binns, R.}
\newblock Fairness and accountability design needs for algorithmic support in
  high-stakes public sector decision-making.
\newblock In {\em Proceedings of the 2018 chi conference on human factors in
  computing systems\/} (2018), pp.~1--14.

\bibitem{wachter2017technically}
{\sc Wachter-Boettcher, S.}
\newblock {\em Technically wrong: Sexist apps, biased algorithms, and other
  threats of toxic tech}.
\newblock WW Norton \& Company, 2017.

\bibitem{wiens2019no}
{\sc Wiens, J., Saria, S., Sendak, M., Ghassemi, M., Liu, V.~X., Doshi-Velez,
  F., Jung, K., Heller, K., Kale, D., Saeed, M., et~al.}
\newblock Do no harm: a roadmap for responsible machine learning for health
  care.
\newblock {\em Nature medicine 25}, 9 (2019), 1337--1340.

\bibitem{winner1980artifacts}
{\sc Winner, L.}
\newblock Do artifacts have politics?
\newblock {\em Daedalus\/} (1980), 121--136.

\bibitem{wong2021timelines}
{\sc Wong, R.~Y., and Nguyen, T.}
\newblock Timelines: A world-building activity for values advocacy.
\newblock In {\em Proceedings of the 2021 CHI Conference on Human Factors in
  Computing Systems\/} (New York, NY, USA, 2021), CHI '21, Association for
  Computing Machinery.

\bibitem{woodruff2018qualitative}
{\sc Woodruff, A., Fox, S.~E., Rousso-Schindler, S., and Warshaw, J.}
\newblock A qualitative exploration of perceptions of algorithmic fairness.
\newblock In {\em Proceedings of the 2018 chi conference on human factors in
  computing systems\/} (2018), pp.~1--14.

\bibitem{zellers2019defending}
{\sc Zellers, R., Holtzman, A., Rashkin, H., Bisk, Y., Farhadi, A., Roesner,
  F., and Choi, Y.}
\newblock Defending against neural fake news.
\newblock In {\em Advances in neural information processing systems\/} (2019),
  pp.~9054--9065.

\bibitem{zeng2018linking}
{\sc Zeng, Y., Lu, E., and Huangfu, C.}
\newblock Linking artificial intelligence principles.
\newblock {\em arXiv preprint arXiv:1812.04814\/} (2018).

\bibitem{zhu2018value}
{\sc Zhu, H., Yu, B., Halfaker, A., and Terveen, L.}
\newblock Value-sensitive algorithm design: Method, case study, and lessons.
\newblock {\em Proceedings of the ACM on Human-Computer Interaction 2}, CSCW
  (2018), 1--23.

\bibitem{zimmer2010but}
{\sc Zimmer, M.}
\newblock {``But the data is already public'': On the ethics of research in
  Facebook}.
\newblock {\em Ethics and information technology 12}, 4 (2010), 313--325.

\bibitem{zoshak2021kant}
{\sc Zoshak, J., and Dew, K.}
\newblock Beyond kant and bentham: How ethical theories are being used in
  artificial moral agents.
\newblock In {\em Proceedings of the 2021 CHI Conference on Human Factors in
  Computing Systems\/} (New York, NY, USA, 2021), CHI '21, Association for
  Computing Machinery.

\end{thebibliography}
\bibliographystyle{acm}

\end{document}